\documentclass[preprint]{aastex}
\usepackage{float}

\newcommand{\sgr}{\mbox{SGR\,J1550-5418~}}
\newcommand{\sgrnos}{\mbox{SGR\,J1550-5418}}

\shorttitle{Broadband Spectroscopy of \sgr Bursts}
\shortauthors{Lin et al.}

\begin{document}

\title{Broadband Spectral Investigations of \sgr Bursts}

\author{Lin Lin\altaffilmark{1}, 
	Ersin G\"o\u{g}\"u\c{s}\altaffilmark{1},
	Matthew G. Baring\altaffilmark{2},
	Jonathan Granot\altaffilmark{3},
	Chryssa Kouveliotou\altaffilmark{4},
	Yuki Kaneko\altaffilmark{1},
	Alexander van der Horst\altaffilmark{5},
	David Gruber\altaffilmark{6},
	Andreas von Kienlin\altaffilmark{6},
	George Younes\altaffilmark{7},
	Anna L. Watts\altaffilmark{5},
	Neil Gehrels\altaffilmark{8}
	}

\email{linlin@sabanciuniv.edu}

\altaffiltext{1}{Sabanc\i~University, Faculty of Engineering and Natural Sciences, Orhanl\i$-$ Tuzla, \.{I}stanbul 34956, Turkey}
\altaffiltext{2}{Department of Physics and Astronomy, Rice University, MS-108, P.O. Box 1892, Houston, TX 77251, USA}
\altaffiltext{3}{The Open University of Israel, 1 University Road, POB 808, Ra'anana 43537, Israel}
\altaffiltext{4}{Space Science Office, VP62, NASA/Marshall Space Flight Center, Huntsville, AL 35812, USA}
\altaffiltext{5}{Astronomical Institute "Anton Pannekoek," University of Amsterdam, Postbus 94249, 1090 GE Amsterdam, The Netherlands}
\altaffiltext{6}{Max-Planck-Institut f\"ur extraterrestrische Physik, Postfach 1312, 85748, Garching bei München, Germany}
\altaffiltext{7}{USRA, National Space Science and Technology Center, 320 Sparkman Dr., Huntsville Al., 35805, USA}
\altaffiltext{8}{NASA Goddard Space Flight Center, Greenbelt, MD 20771, USA}

\begin{abstract}
We present the results of our broadband spectral analysis of 42 \sgr bursts simultaneously detected with the \textit{Swift}/X-ray Telescope (XRT) and the \textit{Fermi}/Gamma-ray Burst Monitor (GBM), during the 2009 January active episode of the source. The unique spectral and temporal capabilities of the XRT Windowed Timing mode have allowed us to extend the GBM spectral coverage for these events down to the X-ray domain ($0.5-10$ keV). Our earlier analysis of the GBM data found that the \sgr burst spectra were described equally well with a Comptonized model or with two blackbody functions; the two models were statistically indistinguishable. Our new broadband (0.5 $-$ 200 keV) spectral fits show that, on average, the burst spectra are better described with two blackbody functions than with the Comptonized model. Thus, our joint XRT/GBM analysis clearly shows for the first time that the \sgr burst spectra might naturally be expected to exhibit a more truly thermalized character, such as a two-blackbody or even a multi-blackbody signal. Using the \textit{Swift} and \textit{RXTE} timing ephemeris for \sgr we construct the  distribution of the XRT burst counts with  spin phase and find that it is not correlated with the persistent X-ray emission pulse phase from \sgrnos. These results indicate that the burst emitting sites on the neutron star need not be co-located with hot spots emitting the bulk of the persistent X-ray emission. Finally, we show that there is a significant pulse phase dependence of the XRT burst counts, likely demonstrating that the surface magnetic field of \sgr is not uniform over the emission zone, since it is anticipated that regions with stronger surface magnetic field could trigger bursts more efficiently.

\end{abstract}

\keywords{pulsars: individual (\sgrnos, 1E\,$1547.0-5408$, PSR\,J$1550-5418$) -- stars: neutron -- X-rays: bursts}

\section{Introduction}

Soft Gamma Repeaters (SGRs) and Anomalous X-ray Pulsars (AXPs) are the observational manifestations of magnetars -  isolated neutron stars possessing extreme magnetic fields, B $>$ $10^{14}$\,G \citep{duncan1992,kouveliotou1998}. Besides being bright X-ray sources, SGRs and AXPs emit intense bursts in hard X-rays / soft $\gamma$-rays on a highly unpredictable frequency with peak luminosities ranging from $10^{38}$ erg s$^{-1}$ to $>$ $10^{47}$ erg s$^{-1}$. These energetic events are attributed to the cracking of the solid neutron star crust by magnetic stress build-up \citep{td95} or to magnetic field line reconnection \citep{lyutikov2003}. For detailed reviews on SGRs and AXPs, see \citet{woods2006} and \citet{mereghetti2008}.

\sgr was discovered as a point source with the \textit{Einstein} Observatory while searching for X-ray emission from radio emitting supernova remnants \citep{lamb1981}. The source was suggested as a magnetar candidate by the similarity of its persistent X-ray spectrum to AXPs and its association with a young supernova remnant \citep{gelfand2007}. Its magnetar nature was confirmed with the detection of radio pulsations with $P=2.096$\,s and $\dot P=2.318\times10^{-11}$, corresponding to an inferred surface dipole magnetic field strength of $2.2\times10^{14}$\,G \citep[AXP\,1E$1547.0-5408$;][]{camilo2007}. An accurate source location was also derived from the radio image, (J2000) R.A.$=15^{\rm h}50^{\rm m}54\fs11 \pm 0\fs01$, decl.$=-54\arcdeg18\arcmin23\farcs7 \pm 0\farcs1$ \citep{camilo2007}. X-ray pulsations at the same spin period were later found with a deeper \textit{XMM-Newton} observation \citep{halpern2008}. No bursts were detected from \sgr until 2008 October, when both the \textit{Swift}/Burst Alert Telescope (BAT) and the \textit{Fermi}/Gamma-ray Burst Monitor (GBM) were triggered by numerous bursts from the source \citep{israel2010,avk2012}. \sgr entered an episode of more active bursting in late 2009 January, and no more burst was detected after 2009 April. During these active episodes, several high energy instruments, such as the \textit{Swift}/BAT and X-Ray Telescope (XRT), the \textit{Fermi}/GBM, the \textit{Rossi X-ray Timing Explorer}(\textit{RXTE})/Proportional Counter Array (PCA), and the \textit{International Gamma-Ray Astrophysics Laboratory} (\textit{INTEGRAL})/Imager on Board the INTEGRAL Satellite (IBIS) and the SPectrometer on INTEGRAL (SPI) recorded hundreds of bursts \citep{mereghetti2009,savchenko2010,kaneko2010,scholz2011,avdh2012,avk2012}. Following these burst active periods, both the persistent X-ray emission characteristics and the spin-down behavior of the source changed remarkably \citep{enoto2010,ng2011,bernardini2011,scholz2011,dib2012,kuiper2012}. 

The spectral properties of \sgr bursts have been extensively studied using {\it individual} instruments: SPI \citep{mereghetti2009}, IBIS \citep{savchenko2010}, BAT \citep{israel2010}, XRT \citep{scholz2011} and GBM \citep{avdh2012}. The XRT data cover a relatively narrow energy range ($0.5-10$\,keV) of the spectrum of a typical SGR burst. Nevertheless, \citet{scholz2011} modeled \sgr burst spectra using the XRT data only in the energy range of $0.5-10$\,keV with a single power law and found an average  photon index of 0.17 $\pm$ 0.33. They also reported that there is a slight anti-correlation between the photon index and the absorbed X-ray flux. BAT provides a spectral energy coverage for SGR bursts from 15\,keV to 150\,keV. \citet{israel2010} found that the spectra of BAT detected bursts can be well described by a single blackbody function with temperatures $\sim 10$\,keV. Finally, in \citet{avdh2012} we derived the spectra for a large set of \sgr bursts detected with GBM in 2009 January using several continuum models. We found that in a slightly broader energy range ($8-200$\,keV), a Comptonization model or the sum of two blackbody functions  (BB$+$BB) can fit equally well the \sgr burst spectra. Note that these two models were also used to describe the spectra of other magnetar bursts in a similar energy range, and revealed intriguing physical insights into the burst phenomena \citep{feroci2004,olive2004,israel2008,lin2011}. 

In this paper, we combine the spectral data of bursts observed simultaneously with XRT and GBM, to investigate their spectral characteristics over a broader energy band ($0.5-200$\,keV). In particular, we concentrate on the two most plausible representations of SGR burst spectra, namely the Comptonization model and the BB$+$BB. Focusing on bursts with data collected over broader spectral bands enhances the chance to discriminate between different spectral models. In Section \ref{sec:obs}, we describe both the XRT and GBM observations of \sgr bursts and the selection of their common events sample. We present the data reduction and analysis in Section \ref{sec:data}. The broad band spectral analysis results and their physical interpretation are presented in Sections \ref{sec:specresult} \& \ref{sec:disc}, respectively.

\begin{deluxetable}{cccc}
\tablecolumns{4}
\tablecaption{XRT observations of \sgr with simultaneous events with GBM. \label{tab:xrt}}
\tablewidth{0pt}
\tablehead{
\colhead{Observation ID} & \colhead{Date} & \colhead{Start Time} & \colhead{Exposure}\tablenotemark{a}\\
\colhead{ } & \colhead{ } & \colhead{  } & \colhead{(ks)}
}
\startdata
00340573000 & 2009 January 22 & 02:26:22 & 6.38 \\
00340573001 & 2009 January 22 & 09:18:28 & 9.45 \\
00030956035 & 2009 January 30 & 17:49:33 & 2.97 \\
\enddata
\tablenotetext{a}{in WT mode}
\end{deluxetable}

\section{Observations and Simultaneous Burst Identification \label{sec:obs}}
\subsection{Observations}

The \textit{Swift}/XRT is an X-ray imaging spectrometer sensitive to photons in the $0.3-10$\,keV energy range \citep{burrows2005}. The telescope is operated either in Photon Counting (PC) mode or in Windowed Timing (WT) mode. Both modes provide the same spectral capabilities, however, the temporal resolution of the PC mode data is about 2.5\,s, too coarse for the very short ($\lesssim100$\,ms) SGR bursts. Therefore, we only employ here XRT observations performed in WT mode, because of its 1.7\,ms readout time well suited to the SGR burst durations. XRT monitored the source with 46 pointed observations in WT mode between 2008 October 01 and 2009 April 30, covering burst active episodes of \sgrnos. These XRT observations were densely concentrated around the most burst active period (2009 January) with a total exposure time of $\sim175$\,ks. Three out of these 46 pointings (listed in Table \ref{tab:xrt}) included bursts simultaneously detected with GBM, as described in detail in the next section.

The \textit{Fermi}/GBM monitors the entire sky (excluding the portion occulted by the Earth) in the energy range from 8 keV to 1 MeV with twelve NaI detectors and in the $0.2-40$\,MeV energy band with two BGO detectors. In its trigger mode, GBM records Time-Tagged Event (TTE) data with high temporal and spectral resolution of $2\,\mu s$ and 128 energy channels, respectively. The trigger readout lasts for 600 s \citep[see][for more details of the instrument and data types]{meegan2009}. 
Using the same burst finding algorithm described in \citet{avdh2012}, we searched for triggered and un-triggered events from \sgr during the active periods of 2008 and 2009\footnote{The list of GBM triggered events is available from http://gammaray.nsstc.nasa.gov/gbm/science/magnetars/magn1550triggers.html}.
In total we identified 692 bursts out of which, 458 events had TTE data. We only used NaI detector TTE data for our spectral analyses, as typical SGR bursts are not detected above 200\,keV, and last only for a fraction of a second.

\subsection{Identification of Simultaneous Events \label{sec:sample}}

\begin{figure}
\includegraphics{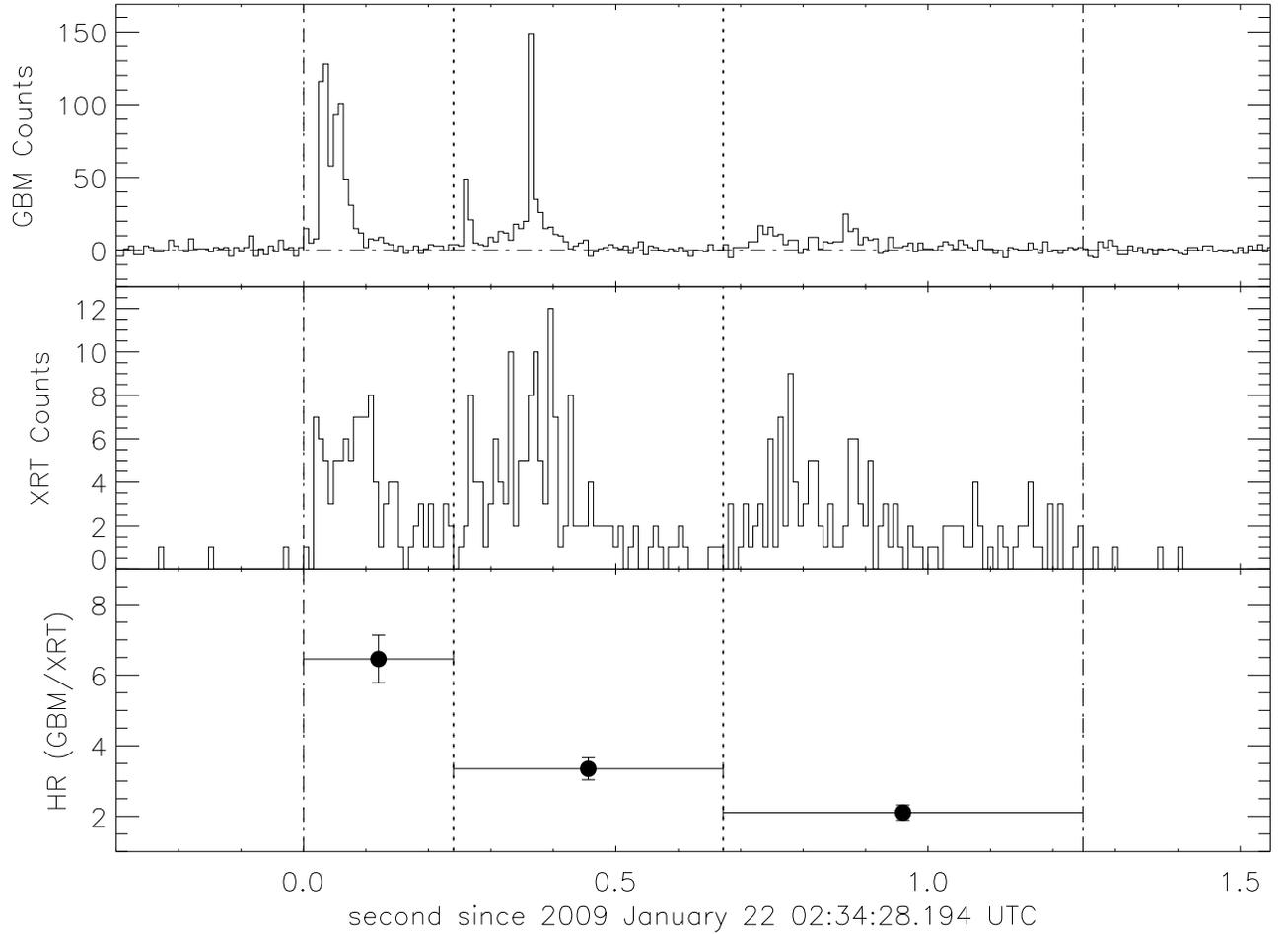}
\caption{ \textit{Top panel}: Background subtracted lightcurve of a burst detected at 02:34:28.194 on 2009 January 22 from \sgr with the GBM NaI-6 detector. It is binned with a time resolution of 8 ms. \textit{Middle panel}: XRT lightcurve of the same burst in 8 ms time resolution. \textit{Bottom panel}: hardness ratios (GBM \textit{v.s.} XRT) of the three sub-intervals indicated with the dotted lines. The dot-dashed lines denote the time interval over which the burst spectrum has been accumulated. \label{fig:lightcurve}}
\end{figure}

To identify the events observed simultaneously with GBM and XRT, we compared the times of the 458 GBM bursts with the time intervals of the XRT observations in WT mode and found 87 common bursts. Note that some SGR bursts have multi-peaked time profiles and each peak was labeled as a burst in our initial un-triggered event search. We used the convention described in \citet{avdh2012} to determine whether multiple peaks constituted a single event, namely, we requested that the time difference between successive burst peaks was less than a quarter of the spin period of \sgr ($\sim0.5$\,s). As a result, we obtained 66 \sgr events simultaneously observed with GBM and XRT. For each burst, we plotted the burst lightcurves seen with GBM and XRT to determine the time interval that includes the main emission episode and is used as our spectral extraction interval (see Figure \ref{fig:lightcurve}). We excluded nine dim bursts that had less than forty counts in the XRT data during the burst interval, statistically not enough for spectral analysis. Additionally, we excluded five very bright bursts which saturated the High Speed Science Data Bus of GBM \citep[see also][]{avdh2012}.

We checked all common bursts in the XRT data for pile-up. We regenerated the level 2 data by including photons in all grades ($0-15$) from level 1 data at the position of \sgrnos, using the standard XRT data processing tasks in HEASOFT. Then we calculated the average readout time for each detection area\footnote{$7\times1$ pixels for XRT WT mode.}. We found that, besides the five GBM saturated events, in ten other bursts the average readout time for central detection areas is smaller than 1.7\,ms, the smallest readout time for the WT mode. This indicates that these bursts are affected by the photon pile-up in the XRT data. Compared to the remaining common events, these bursts have more photons in grades higher than 2; as a result an analysis of the data in the good grade range ($0-2$) would lack most information and the results would be misleading. Therefore, we also excluded these ten piled-up bursts from further investigations. The final outcome of all these filters was a selection of 42 bursts, observed with both GBM and XRT, that we then used for broadband spectral analysis (40 bursts detected on 2009 January 22 and 2 events on 2009 January 30).

\subsection{Data Reduction \label{sec:data}}

We extracted time integrated spectra for our 42 simultaneously detected bursts. In Figure \ref{fig:lightcurve}, we present the burst detected at 02:34:28.194 on 2009 January 22 as an illustrative example of spectral integration ranges. We describe below the procedures we followed to generate the GBM and XRT spectra.

For the GBM data, we selected the NaI detectors with a source angle smaller than $60\arcdeg$, and without any blockage from the Large Area Telescope (LAT) or other parts of the satellite, such as solar panels and radiators. We determined background levels by fitting pre and post burst intervals with a first order polynomial using \textit{RMFIT v3.4rc12}\footnote{R.S. Mallozzi, R.D. Preece, \& M.S. Briggs, "RMFIT, A Lightcurve and Spectral Analysis Tool," \copyright 2008 Robert D. Preece, University of Alabama in Huntsville, 2008} and extracted both burst and background spectra. We then used \textit{grppha} to group the extracted source spectrum to include at least 15 source counts in each energy bin. We generate the response matrices using \textit{GBMRSP v1.9} for each burst. 

For the XRT data, we selected events with grade range of $0-2$ with \textit{xselect}, and accumulated the source spectra from a 40 pixel long section of the chip centered at the \sgr location in the same time intervals as used for the GBM spectra. We extracted background spectra from a region of the same size, away from the source. We generated the Ancillary Response Function (ARF) file for each burst using \textit{xrtmkarf} in HEASOFT. In our spectral fitting we used the standard response file `\textit{swxwt0to2s6\_20070901v012.rmf}' provided in the \textit{Swift} calibration database. Finally, we also grouped the source spectra to include a minimum of 15 counts in each energy bin. We fit all spectra using XSPEC v12.7.0. 

\section{Spectral Analysis \label{sec:specresult}}

\begin{figure*}
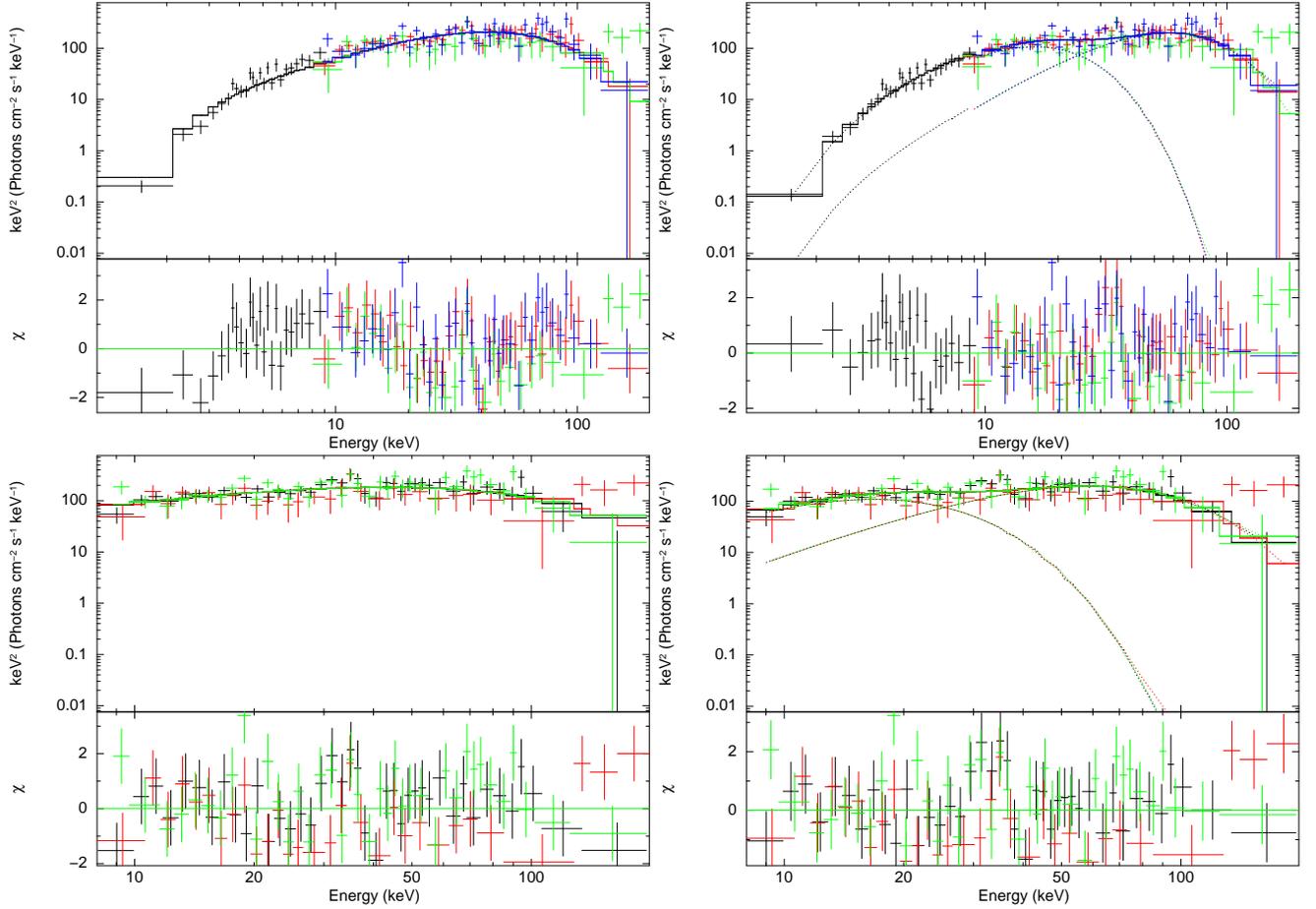

\includegraphics*[scale=0.35, angle=270]{01compt.eps}
\includegraphics*[ scale=0.35, angle=270]{01bbbb.eps}\\
\includegraphics*[scale=0.35, angle=270]{01compt_gbm.eps}
\includegraphics*[scale=0.35, angle=270]{01bbbb_gbm.eps}
\caption{The spectrum of the \sgr burst shown in Figure \ref{fig:lightcurve}. Top two panels show the XRT-GBM joint fit spectrum. Bottom two panels show the GBM only spectrum. The left column are COMPT model fits and the right column are fits with a BB$+$BB model. The lower parts in each panel show the fit residuals.  \label{fig:spec}}
\end{figure*}

Motivated by our recent results published in \citet{lin2011} and \citet{avdh2012}, we modeled the broadband time-integrated spectra of all 42 common events with the Comptonized model (COMPT) and the sum of two blackbody functions (BB$+$BB). The COMPT model is in a single power law shape with a high energy exponential cutoff expressed as:
\begin{displaymath}
f = A \exp[-E(2+\lambda)/E_{\rm{peak}}] (E/E_{\rm{piv}})^\lambda,
\end{displaymath}
where $f$ is the photon flux in photons s$^{-1}$ cm$^{-2}$ keV$^{-1}$, $A$ is the amplitude with units same as $f$, $E_{\rm{peak}}$ is the energy (in keV) at which the spectral distribution function peaks, $\lambda$ is the photon index, and $E_{\rm{piv}}$ is the pivot energy fixed at $20$ keV. The latter BB$+$BB model has been commonly used in the context of SGR burst spectra \citep{feroci2004,olive2004,israel2008}. In our spectral fits, we fix the multiplicative interstellar absorption term at $3.24\times10^{22}$\,cm$^{-2}$, since \sgr bursts are short and XRT burst spectra cannot constrain the absorption parameter\footnotemark \footnotetext{We adopt here the value obtained with the XRT observations of the persistent emission during the burst active episode of \sgr by \citet{scholz2011}}. In our joint fits, we also include a multiplicative factor to account for the cross-calibration between the XRT and GBM spectra. In Figure \ref{fig:spec} we show the broadband spectral modeling results of the burst of Figure \ref{fig:lightcurve} with the COMPT and BB$+$BB models. We notice that a BB$+$BB model fit to the combined XRT and GBM data results in better residuals than the COMPT ones and are also better than the COMPT fits of the GBM data alone (see the two left panels in Figure \ref{fig:spec}). These results indicate that the broadband joint fits can constrain better the physical emission model in SGR bursts as we discuss in Section \ref{sec:joint-gbm}. We calculated the average reduced $\chi^2$ for both models, and obtained 1.03 the BB$+$BB model and 1.11 for the COMPT model. The standard deviation of the reduced $\chi^2$ is 0.20 and 0.26 for the BB$+$BB and the COMPT model, respectively. These results further support the BB$+$BB model against COMPT. 

Table \ref{tab:gbm-xrt} presents the joint fit results with $1\sigma$ errors for all 42 bursts. Columns $1-3$ are the burst numbers, start times in UTC, and durations of the time-integrated spectra. The COMPT model parameters, i.e., the power law index ($\lambda$), $E_{\rm peak}$, and fit statistics are shown in columns $4-6$. Columns $7-13$ correspond to the temperatures of the two blackbody components, the luminosity and size of the emitting area for each blackbody component (assuming a distance to \sgr of 5\,kpc), and the fit statistics of the BB$+$BB model fits. Since the energy flux values obtained with the COMPT model are in agreement with those we get from the BB$+$BB model, we only present in columns 14 \& 15 the energy flux in the GBM energy band ($8-200$\,keV) and the observed flux in the XRT energy band ($0.5-10$\,keV) using the BB$+$BB fits. 


The COMPT model has one less parameter than the BB$+$BB function. Since these two models are not nested, we cannot employ a $\Delta\chi^2$ test to determine which model provides a more adequate fit. Therefore, we performed extensive simulations for each of the 42 bursts. We first fit each joint spectra with the seed model (i.e., the model with the smallest resulting $\chi^2$ value): these were the BB$+$BB model for 33 bursts and the COMPT for the remaining nine bursts). Then, we generated 10000 simulated spectra based on the seed model of each burst and fitted the simulated spectra with both models. For each set of 10000 simulations, we selected fits with well constrained model parameters, requiring the errors of $E_{\rm peak}$ of COMPT and of the two blackbody temperatures to be less than $20\%$. We constructed the distributions of all fit parameters as well as of the fit statistics and fitted them with a Gaussian function. The mean values of these distributions for the seed model parameters agree with the fit results listed in Table \ref{tab:gbm-xrt}. Finally, we calculated the percentage of the simulated spectra which result in a smaller $\chi^2$ value when fitted with the seed model. This percentage, defined as the $p-$value, reflects the significance of the preference of the seed model at a given background and fluctuation level. We conclude that the seed model provides a significantly better fit than the other model if $p > 0.9$. The $p-$values of all bursts are listed in the last column of Table \ref{tab:gbm-xrt}.

\begin{figure*}
\includegraphics*[scale=1]{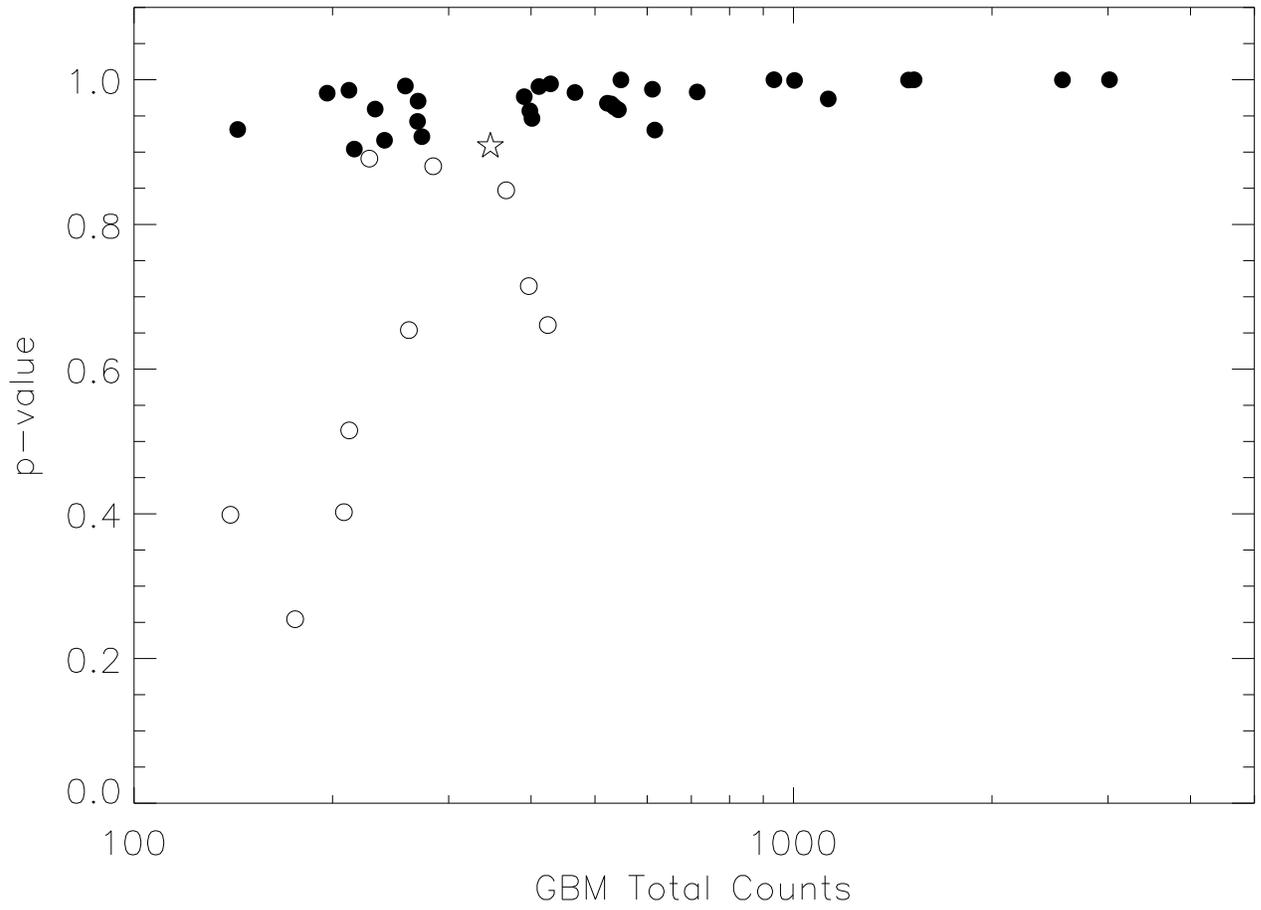}
\caption{Plot of the $p-$value {\it v.s.} the total counts in the GBM energy band ($8-200$\,keV). The black dots are the BB$+$BB bursts, the open circles are the intermediate group bursts and the five point star is the COMPT burst.  \label{fig:p-gbmcnt}}
\end{figure*}

We then grouped the 42 bursts into three categories based on their resulting $p-$values: the BB$+$BB burst, BB$+$BB is significantly prefered to COMPT model; the COMPT burst, COMPT model is significantly better than BB$+$BB; and the intermediate group containing bursts for which both the BB+BB and COMPT models provide equally acceptable fit results ($p < 0.9$). We find that 31 events are the BB$+$BB bursts, only one is a COMPT burst (event $\# 18$), and 10 are in the intermediate group. Figure \ref{fig:p-gbmcnt} displays the $p-$values of all bursts {\it versus} their total GBM counts. Note that a similar trend is obtained if the $p-$values were plotted with their corresponding total XRT counts. We find that the bright bursts prefer the BB$+$BB model to the COMPT model. This might be an indication of higher opacity, on average, in the more luminous bursts. We explore both the statistical and the correlative behavior of all model parameters in the next sections.



\begin{figure*}
\includegraphics*[scale=0.5]{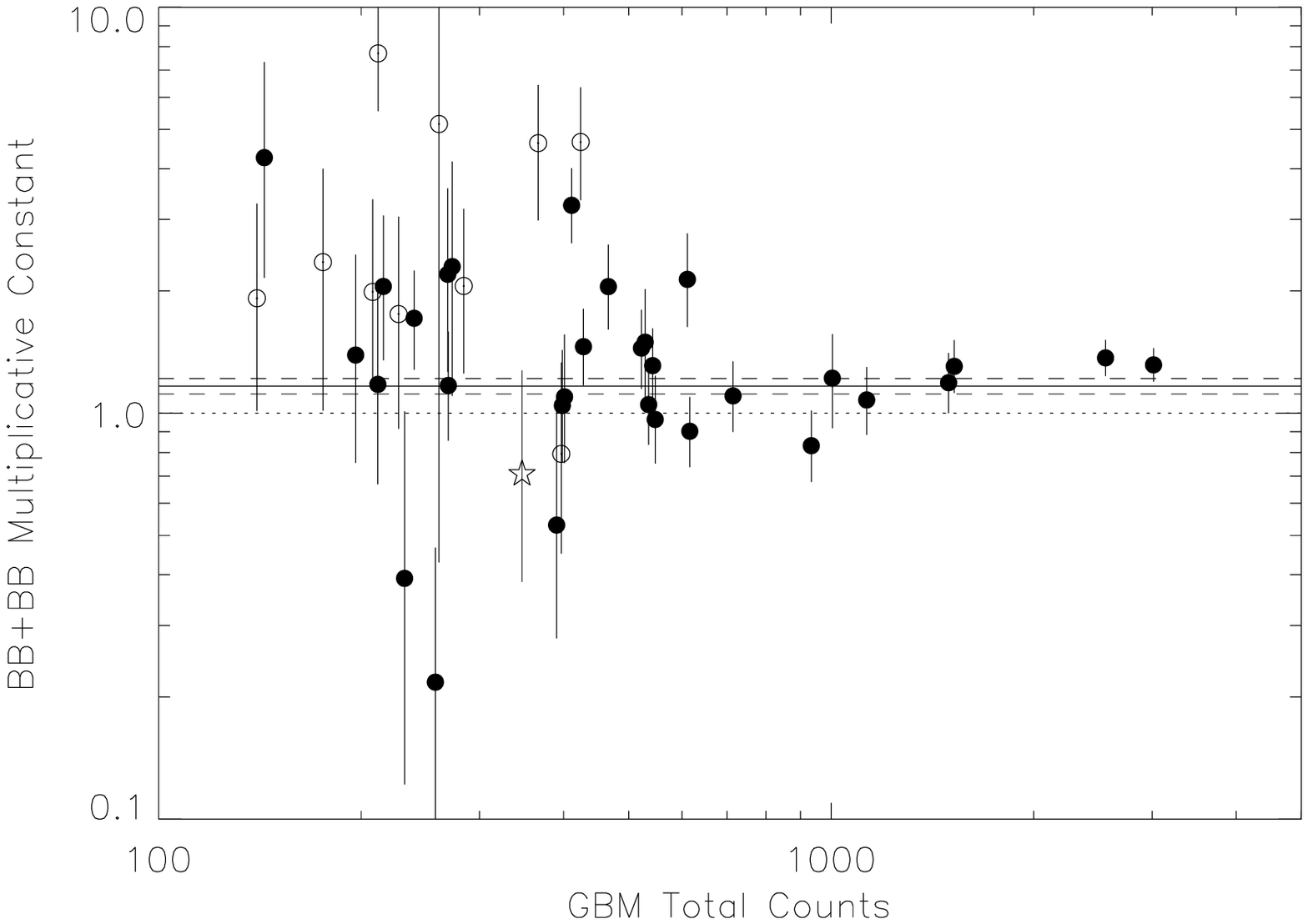}
\includegraphics*[scale=0.5]{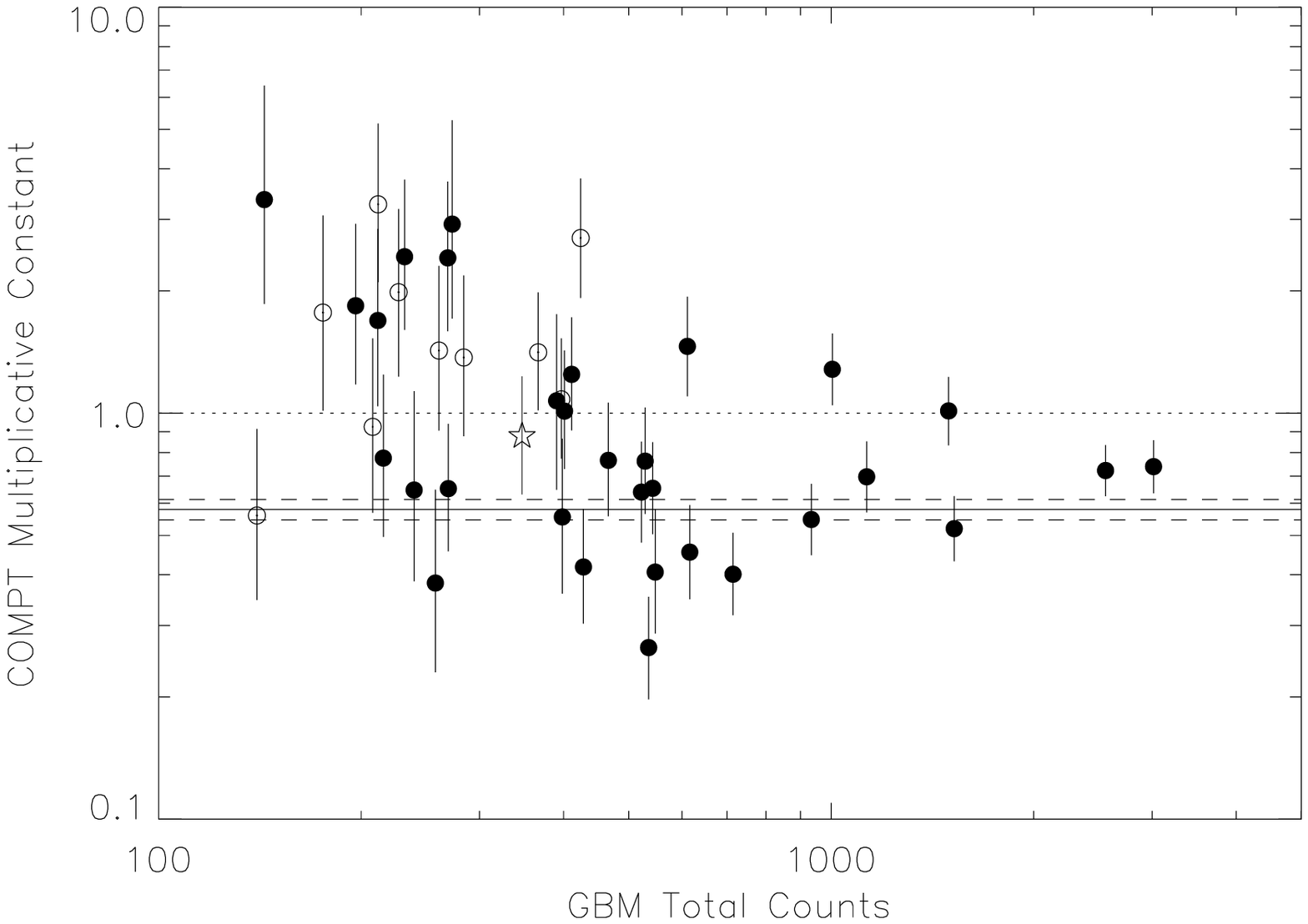}
\caption{Plot of the multiplicative factor values from the BB$+$BB model (\textit{left}) and the COMPT model (\textit{right}) fits  \textit{v.s.} the total counts ($8-200$\,keV) in the brightest GBM detector. The black dots are the BB$+$BB bursts, the open circles are the intermediate group bursts and the five point star is the COMPT burst.The dotted line indicates the factor being equal to 1, and the solid and dashed lines display the weighted average of the factor and its $1\sigma$ error, respectively. \label{fig:const}}
\end{figure*}

Our joint spectral fits provided the opportunity to investigate the cross-calibration of the XRT and GBM instruments with the two spectral models. We determined a multiplicative factor between the XRT and GBM detectors for each model. The left panel of Figure \ref{fig:const} shows the behavior of this factor from the BB$+$BB model fits as a function of total counts in the brightest GBM detector. The same plot for the COMPT model fits is presented in the right panel of Figure \ref{fig:const}. The instruments are perfectly cross-calibrated if the multiplicative factor is equal to 1 (the dotted lines in Figure \ref{fig:const}). For brighter bursts, the factor from the BB$+$BB model fit is better constrained, and it does not change significantly from burst to burst: its weighted mean value and $1\sigma$ error are $1.17 \pm 0.05$. However, the weighted mean of the COMPT model derived factor is $0.58 \pm 0.03$, much smaller and further from 1. We notice that the values obtained from a better fit model are closer to the perfect cross-calibration factor of 1. We find that the constant factor is not significantly correlated with any spectral parameters. We conclude that the cross-calibration of the XRT and GBM instruments works reasonably well within the fluence and energy range of the bursts in our sample. Future efforts to better understand the cross-calibration of these two instruments should include bursts with wider energy and fluence ranges, and perhaps, different spectral models.

\subsection{Comptonized Model \label{sec:compt}}

\begin{figure*}
\includegraphics*[scale=0.5]{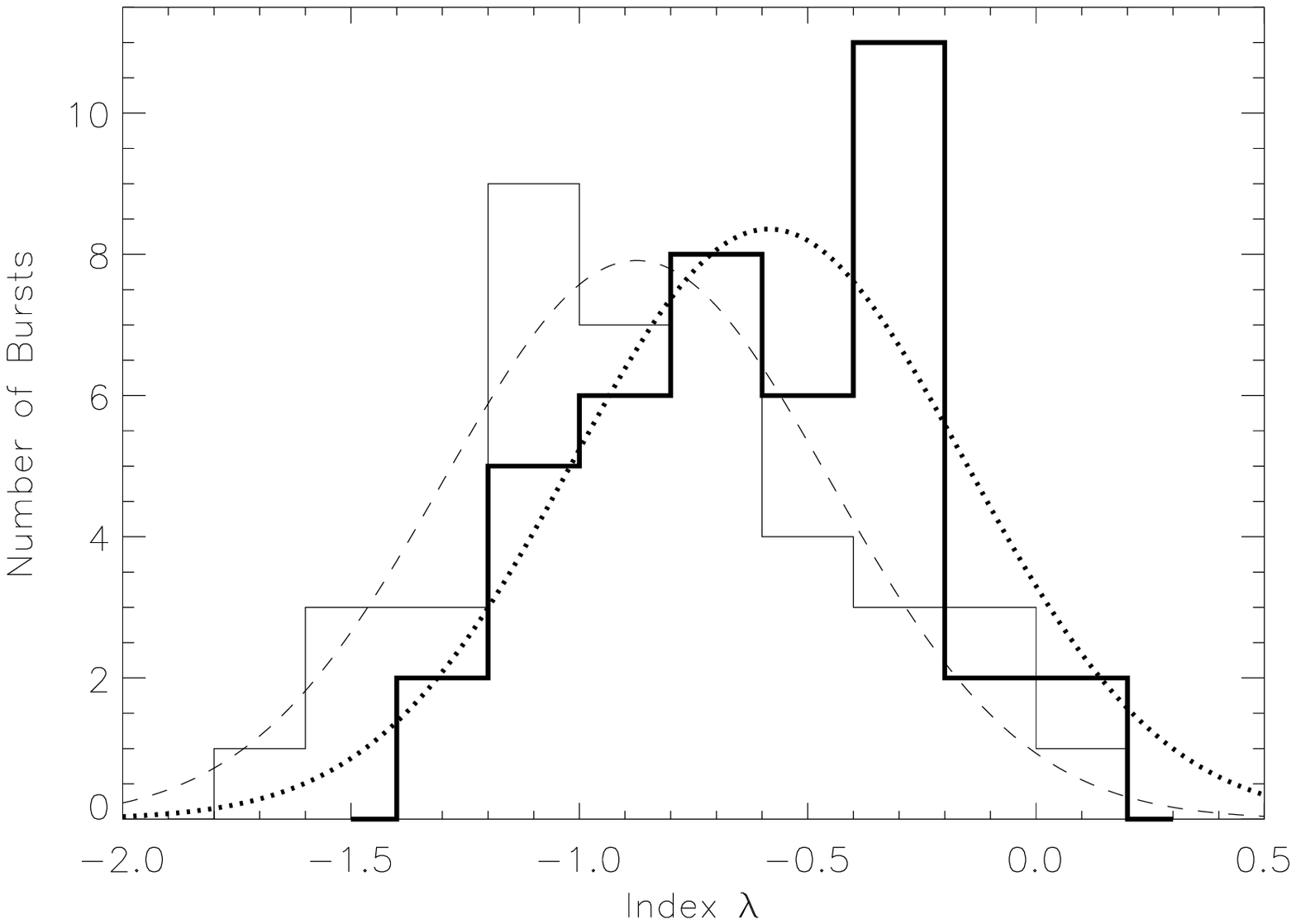}
\includegraphics*[scale=0.5]{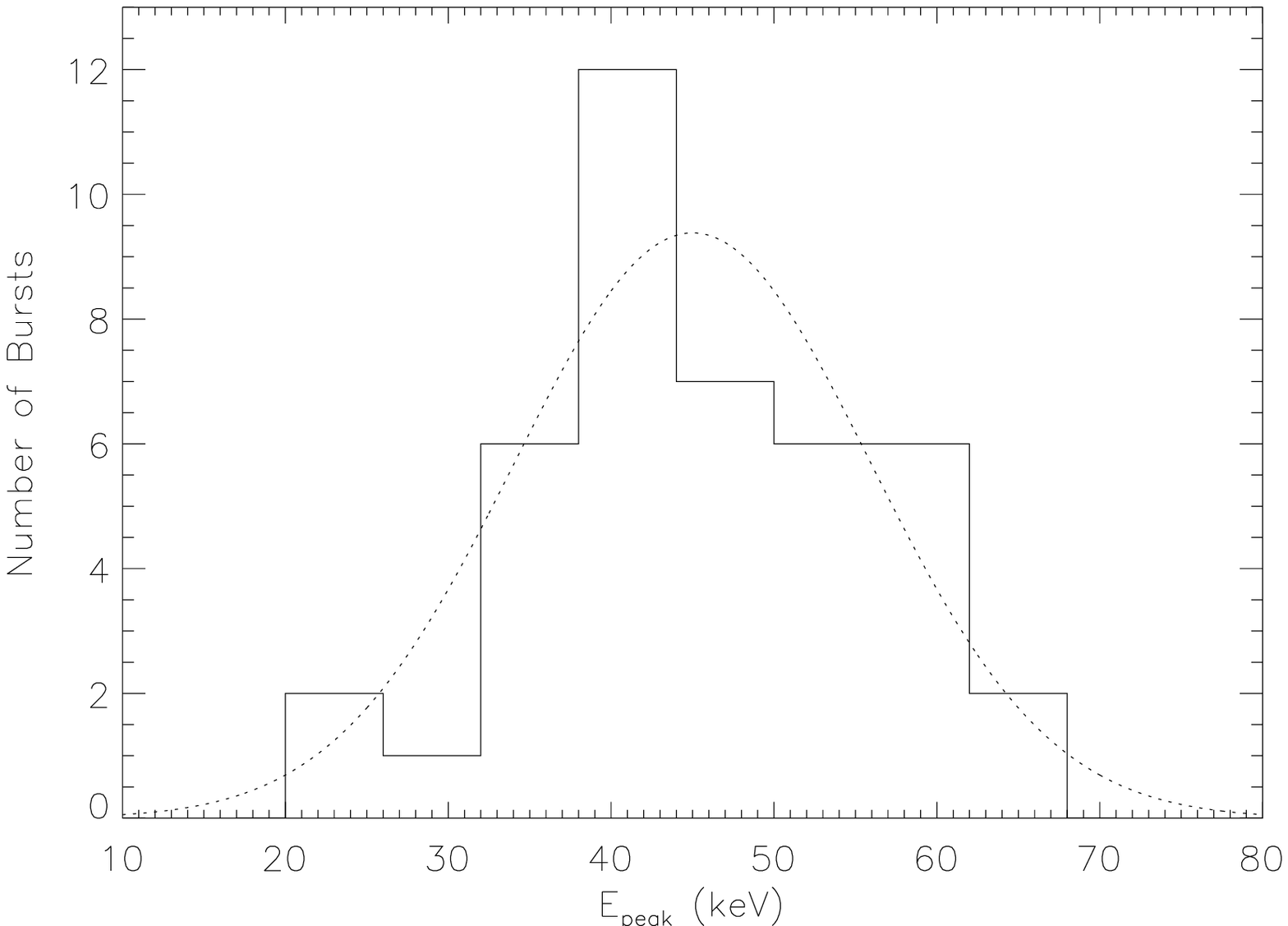}
\caption{Distributions of the COMPT model index (\textit{left}) and $E_{\rm peak}$ (\textit{right}). The dotted lines show the best fit normal functions. The thick histograms in the left panel are the index distribution of the XRT-GBM joint fit, while the thinner histograms are the same distribution for the GBM data only fit. \label{fig:comptpar}}
\end{figure*}

We present the distribution of power law indices obtained from the joint fit to the XRT and GBM spectra (histograms with thick lines) in the left panel of Figure \ref{fig:comptpar}. We also present in the same Figure the distribution of the indices as obtained by fitting the GBM spectra only (histograms with thin lines). It is clear that the latter fits yield, on average, lower power law indices (i.e., the spectrum is harder),  reflecting the overall broadband curvature of the burst spectra. We fit each distribution with a normal function and find that the joint broadband fit mean index value is  $-0.58 \pm 0.09$ (width of $0.43 \pm 0.11$, dotted lines in Figure \ref{fig:comptpar}, left panel), while the  GBM only fit mean index is $-0.87 \pm 0.05$ (width of $0.42 \pm 0.06$, dashed lines in Figure \ref{fig:comptpar}, left panel). We show the distribution of the $E_{\rm peak}$ values obtained from the joint fit in the right panel of Figure \ref{fig:comptpar}. A normal function fit to this distribution yields a mean of $45.0 \pm 2.1$\,keV with a width of $10.9 \pm 2.2$\,keV. The $E_{\rm peak}$ values from joint fit agree with those from GBM data only fit very well, see detailed discussion in Section \ref{sec:joint-gbm}.

\begin{figure*}
\includegraphics*[scale=0.5]{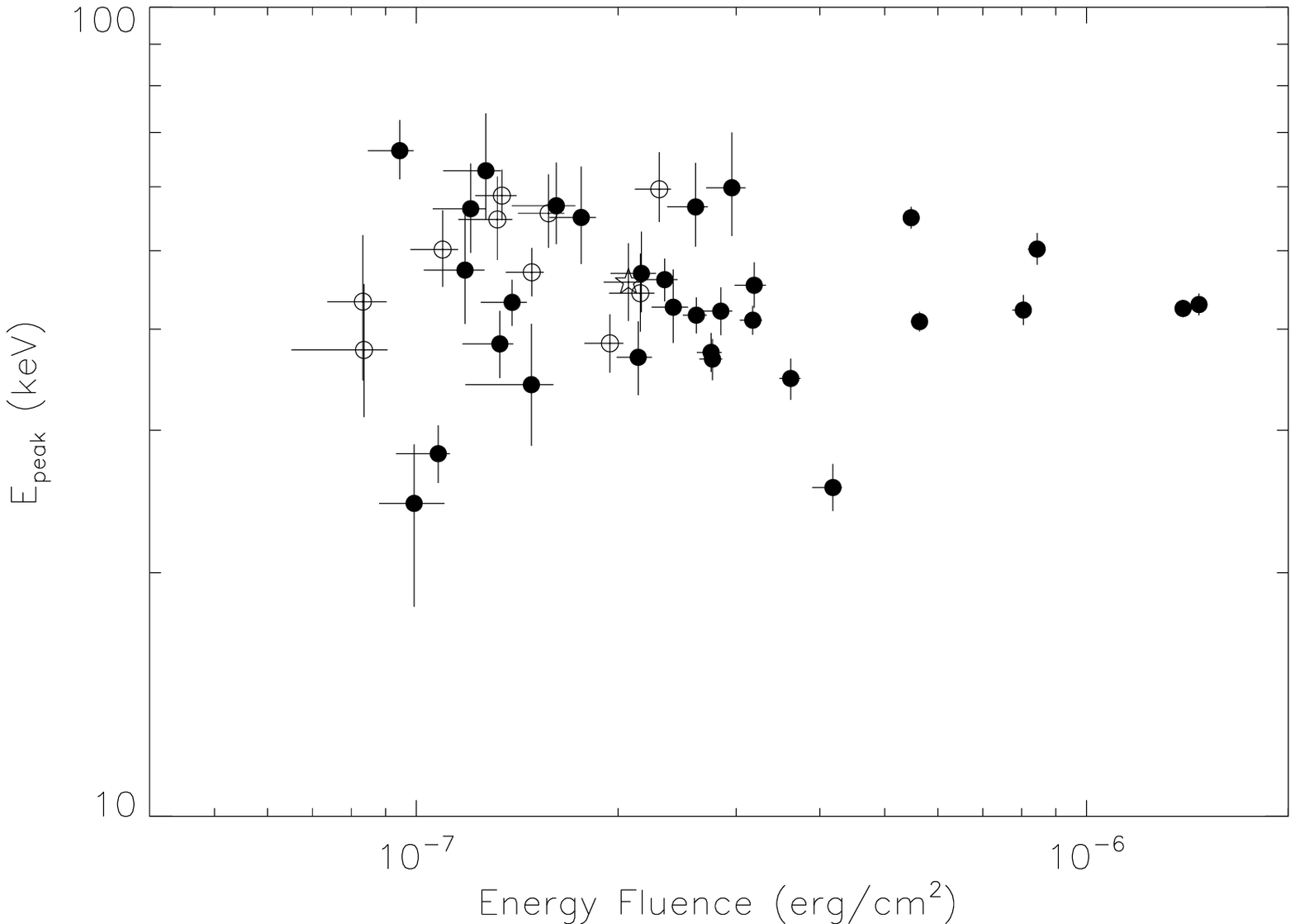}
\includegraphics*[scale=0.5]{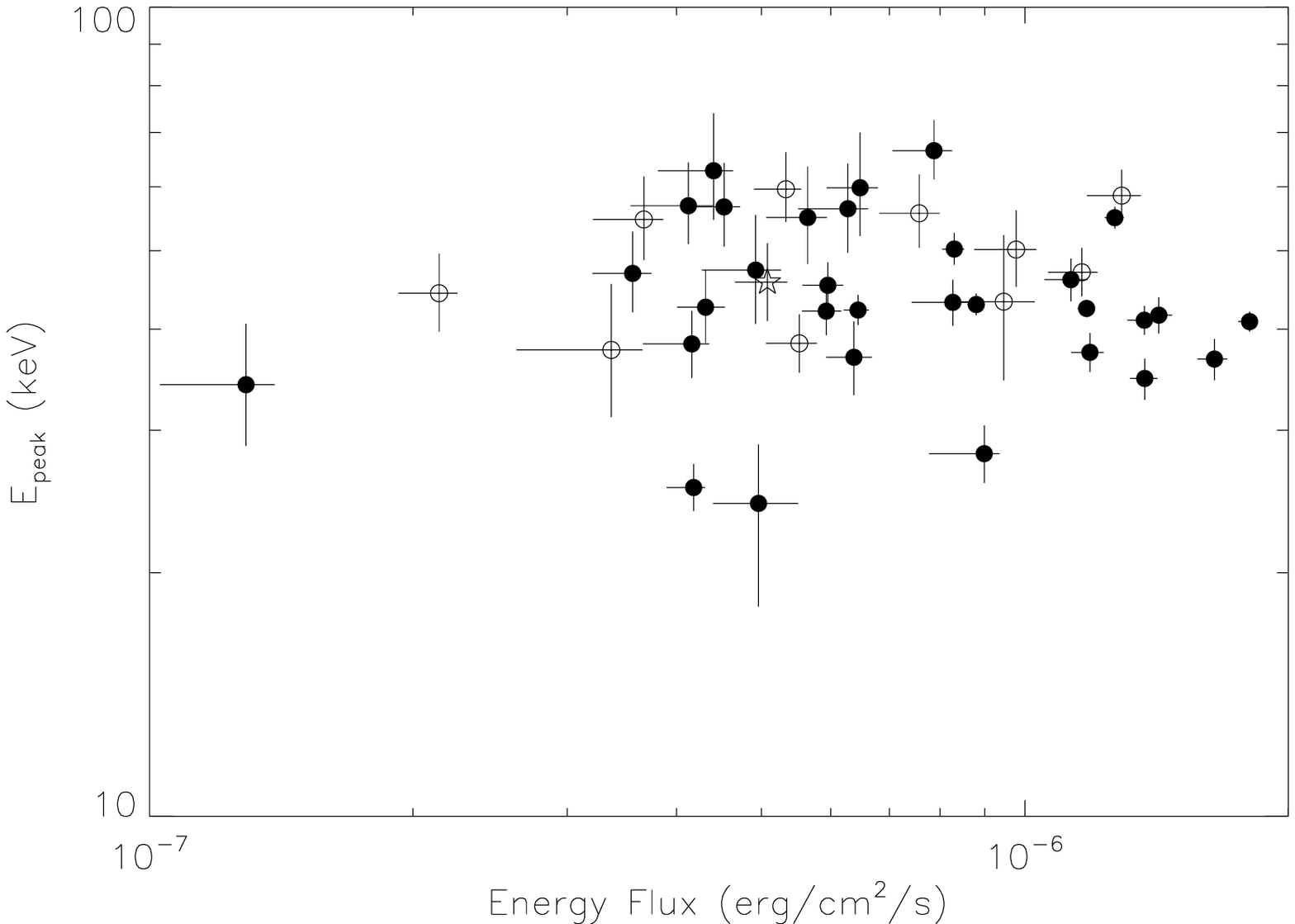}
\caption{Plot of $E_{\rm peak}$ \textit{v.s.} observed energy fluence (\textit{left}) and average flux (\textit{right}) between $0.5-200$\,keV. The black dots are the BB$+$BB bursts, the open circles are the intermediate group bursts and the five point star is the COMPT burst. \label{fig:ep-gbmflnc}}
\end{figure*}

In Figure \ref{fig:ep-gbmflnc}, we present a plot of the joint fits $E_{\rm peak}$ values {\it v.s.} the observed fluence/flux of bursts in the $0.5-200$ keV range. We do not find any anti-correlation or a broken power law trend as seen by fitting the GBM data of \sgr \citep{avdh2012} and SGR 0501+4516 \citep{lin2011}, which we attribute to the fact that the 42 common events cover a much narrower fluence range, about $1/3$ of that in the complete GBM burst sample from the same active period \citep{avdh2012}. 

\subsection{Two Blackbody Model \label{sec:bbbb}}

\begin{figure*}
\includegraphics*[scale=0.5]{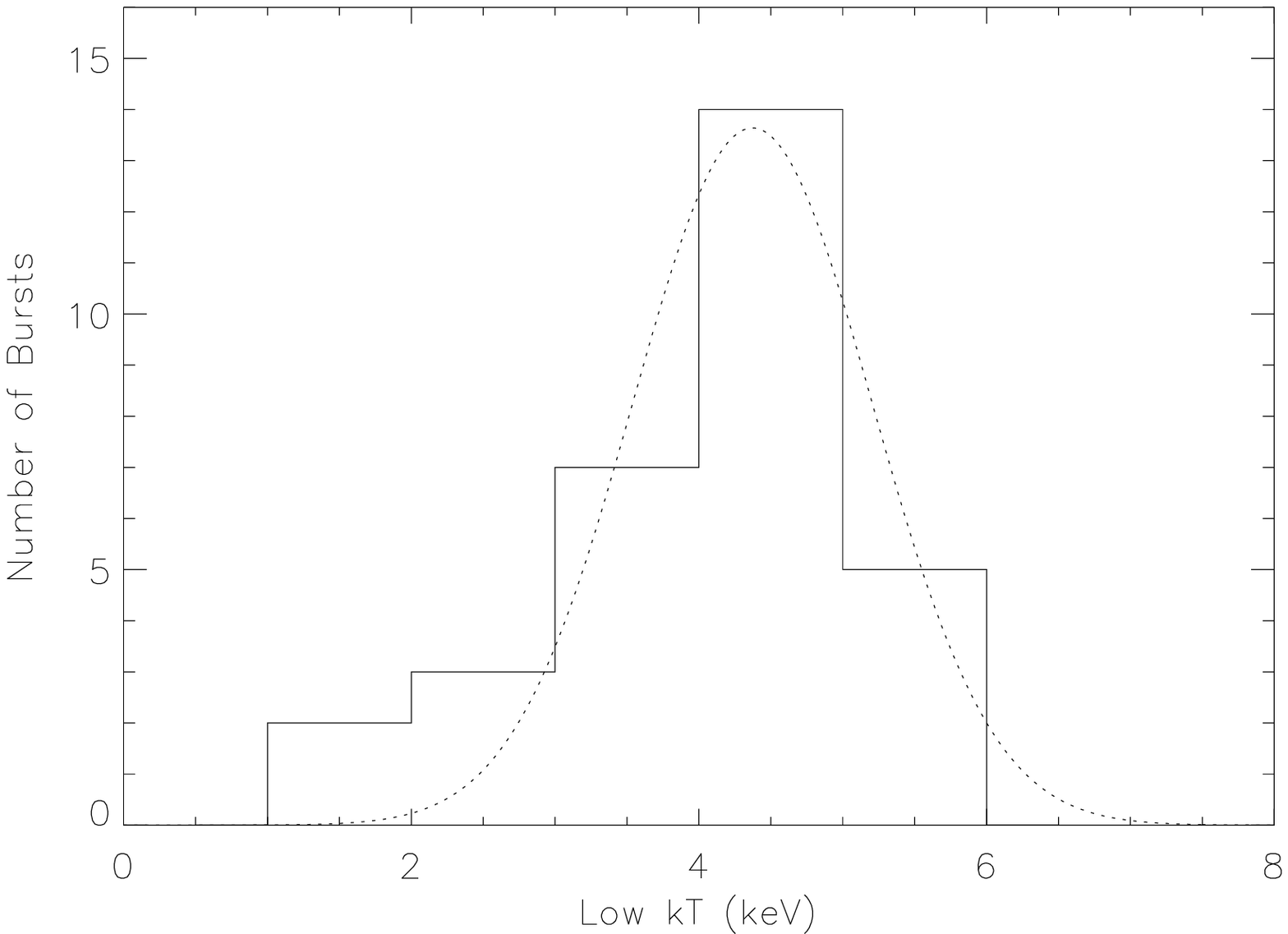}
\includegraphics*[scale=0.5]{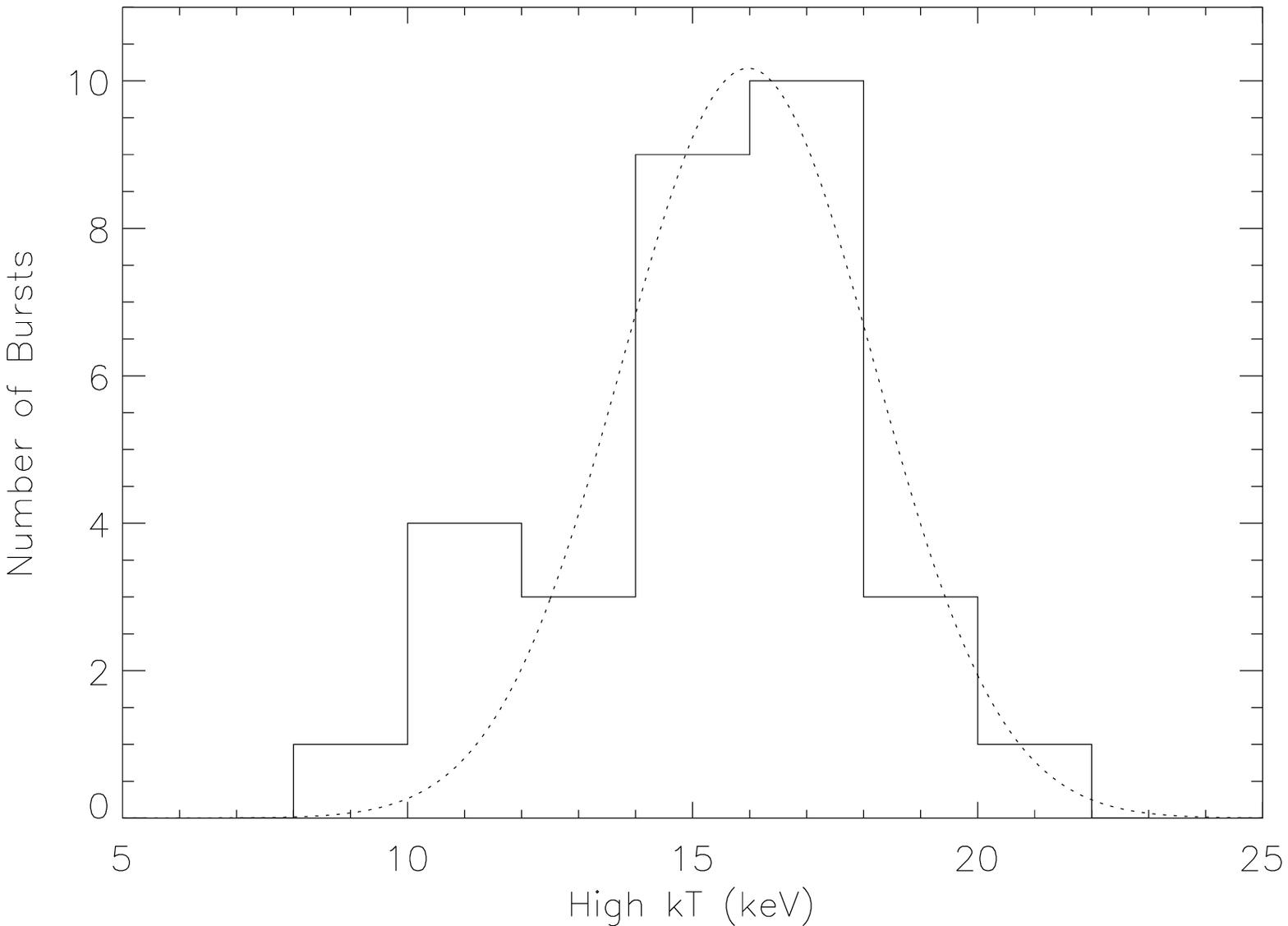}
\caption{Distributions of the temperatures of the two blackbody components of the BB$+$BB model. The dotted lines show the best fit normal functions. \label{fig:bbbbkt}}
\end{figure*}

From simulation, we find that the  BB$+$BB model provides a significantly better fit to 31 common events. We use these BB$+$BB bursts to investigate the properties of the model parameters. In Figure \ref{fig:bbbbkt} we present the distribution of temperatures of the cool (left panel) and hot (right panel) BB components, respectively. We fit a normal function to these distributions and obtain a mean value for the cool BB temperatures of $4.4 \pm 0.2$\,keV (width, $0.8 \pm 0.1$\,keV), and for the hot BB temperatures of $16.0 \pm 0.4$\,keV (width $2.2 \pm 0.4$\,keV). We also calculated the weighted mean values for cool and hot BB temperatures, which are $4.2 \pm 0.1$\,keV and $14.8 \pm 0.2$\,keV, as well as their standerd deviations, 0.9\,keV and 2.7\,keV. These mean temperatures are in agreement within uncertainties with the values obtained by fitting only GBM spectra of \sgr bursts \citep{avdh2012}, and similar to the BB$+$BB temperature values obtained for the bursts of other SGR sources \citep{feroci2004,olive2004,israel2008,lin2011}.


\begin{figure}
\includegraphics{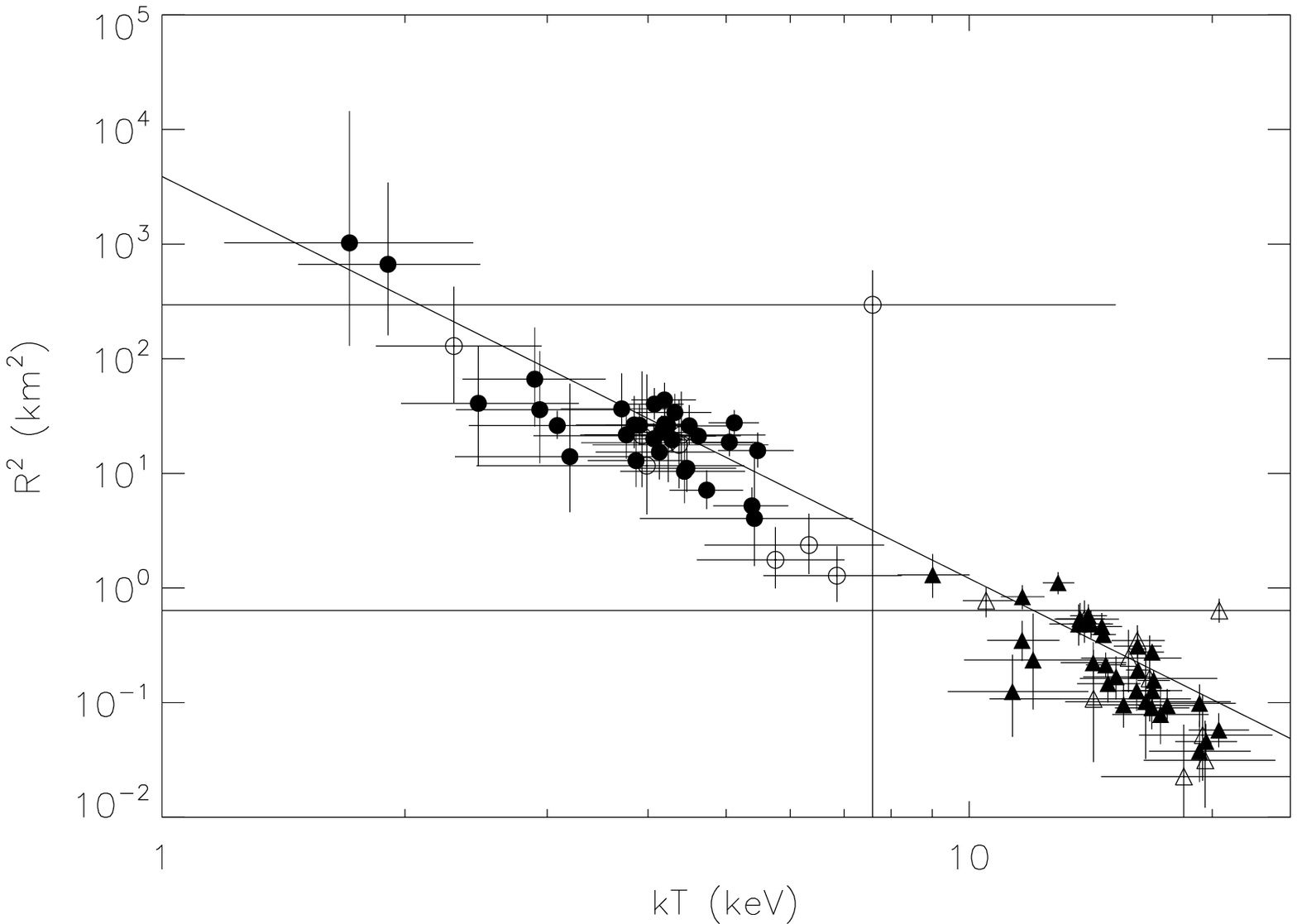}
\caption{Emission area as a function of blackbody temperature for both blackbody components of the BB$+$BB model fit. The hot and cool blackbody components are displayed with triangles and circles, respectively. The filled symbols are the BB$+$BB bursts. The intermediate group bursts are shown as open symbols. The solid line indicate the $R^2 \propto (kT)^{-3.5}$ relation, the best fit power law function with all emission areas and temperatures for BB$+$BB group bursts. \label{fig:r2-kt}}
\end{figure}

\begin{figure}
\includegraphics{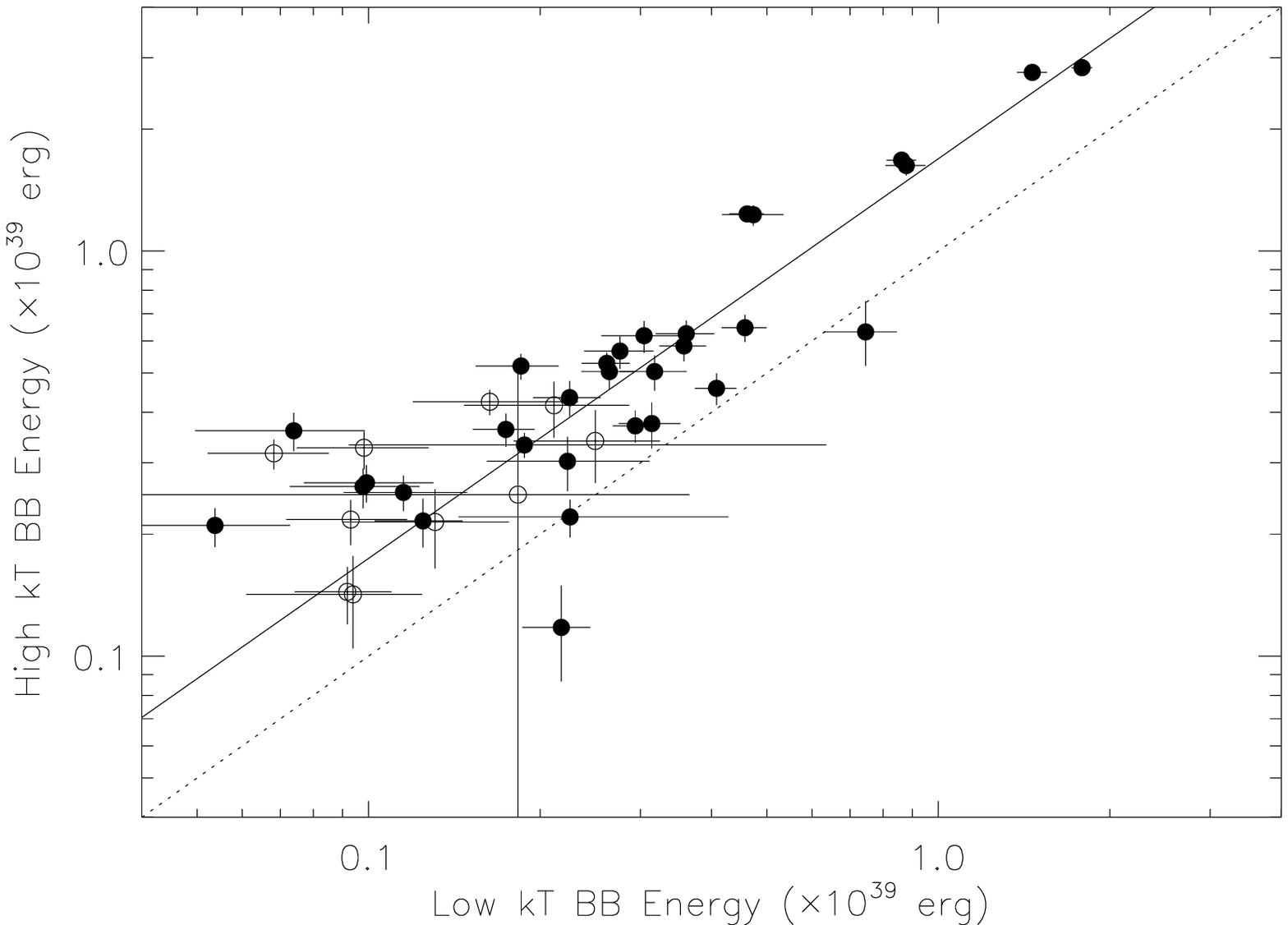}
\caption{Correlation between the total energy emitted from the hot and cool blackbody components. The filled and open circles present the BB$+$BB and intermediate group bursts, respectively. The best power law fit ($E_{\rm hot} \propto E_{\rm cool}^{0.99}$) for the BB$+$BB group is shown as a solid line. The dotted line is where the two BB components have equal energy. \label{fig:he-le}}
\end{figure}

In Figure \ref{fig:r2-kt} we demonstrate the anti-correlation between the emission area and the temperature of the cool and hot BB components. The Spearman Rank Order Correlation test yields a correlation coefficient for the hot BB of $-0.79$ with a chance probability of $1.40 \times 10^{-7}$. The correlation for the cool BB is not as significant as the hot component: its Spearman correlation coefficient is $-0.62$ with a probability of $1.75\times10^{-4}$. We fit the emission area {\it v.s.} temperature for the cool and hot BB with power laws, and obtain power law indices of $-1.5 \pm 1.4$ and $-4.5 \pm 0.9$, for the cool and hot components, respectively. We also fit the emitting area  {\it v.s.} the two BB temperatures together with a single power law and obtain the best fit power law index of $-3.5 \pm 0.2$ (shown as the solid line in Figure \ref{fig:r2-kt}). Note that this value is very close to the theoretical expectation from a single BB with fixed luminosity, $R^2 \propto (kT)^{-4}$. The observed departure at the high temperature end from this ideal form reflects the higher luminosity present in the hot BB component relative to the cooler one; on the average, the hot BB energy is about twice the one emitted from the cool BB. The two energies are highly correlated with a Spearman correlation coefficient of $0.88$, corresponding to a chance probability of $6.45\times10^{-11}$; a power law fit yields an index of $0.99 \pm 0.05$ (Figure \ref{fig:he-le}).

\section{Discussion \label{sec:disc}}
\subsection{Impact of the XRT data \label{sec:joint-gbm}}

\begin{figure*}
\includegraphics*[scale=0.4]{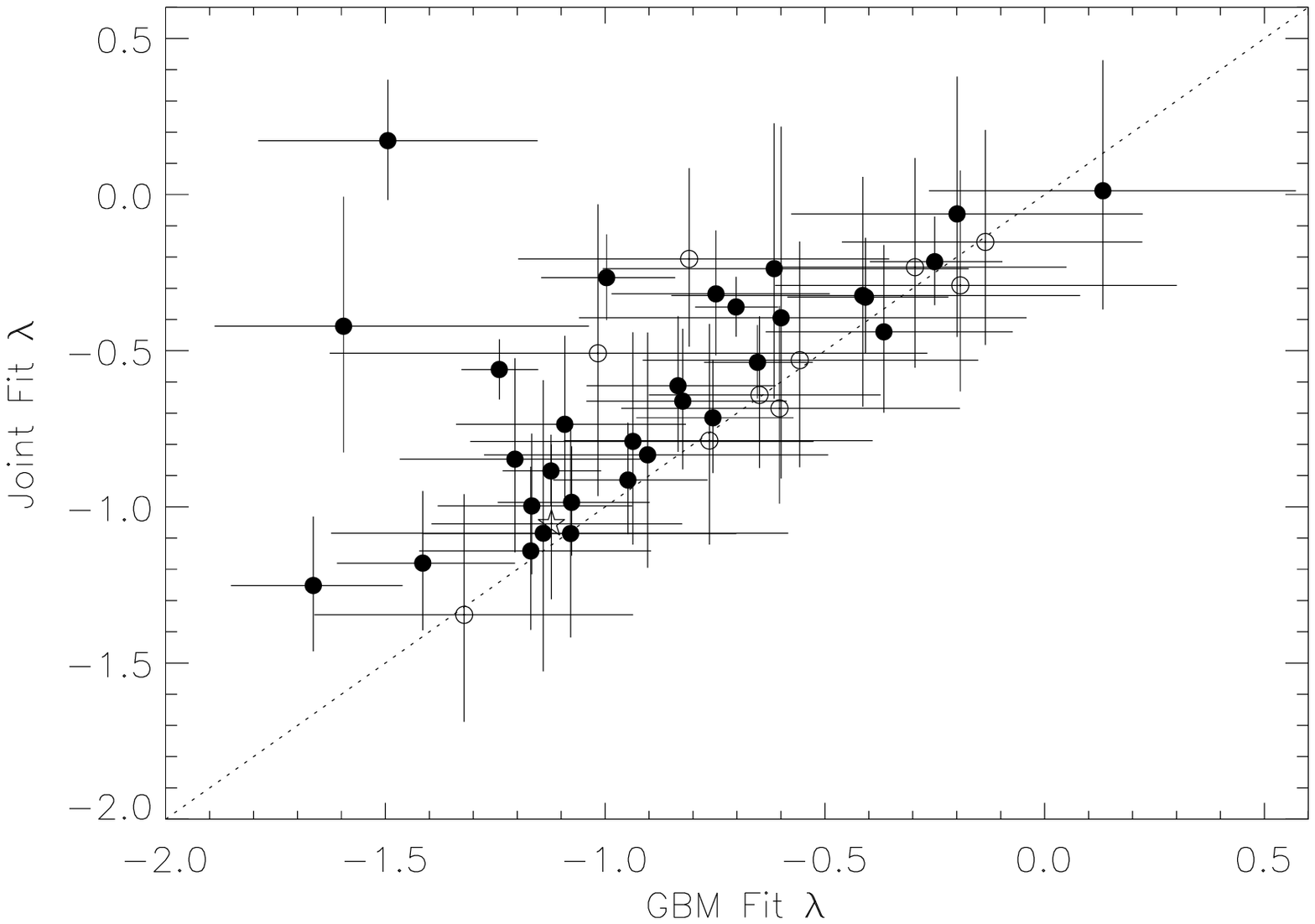}
\includegraphics*[scale=0.4]{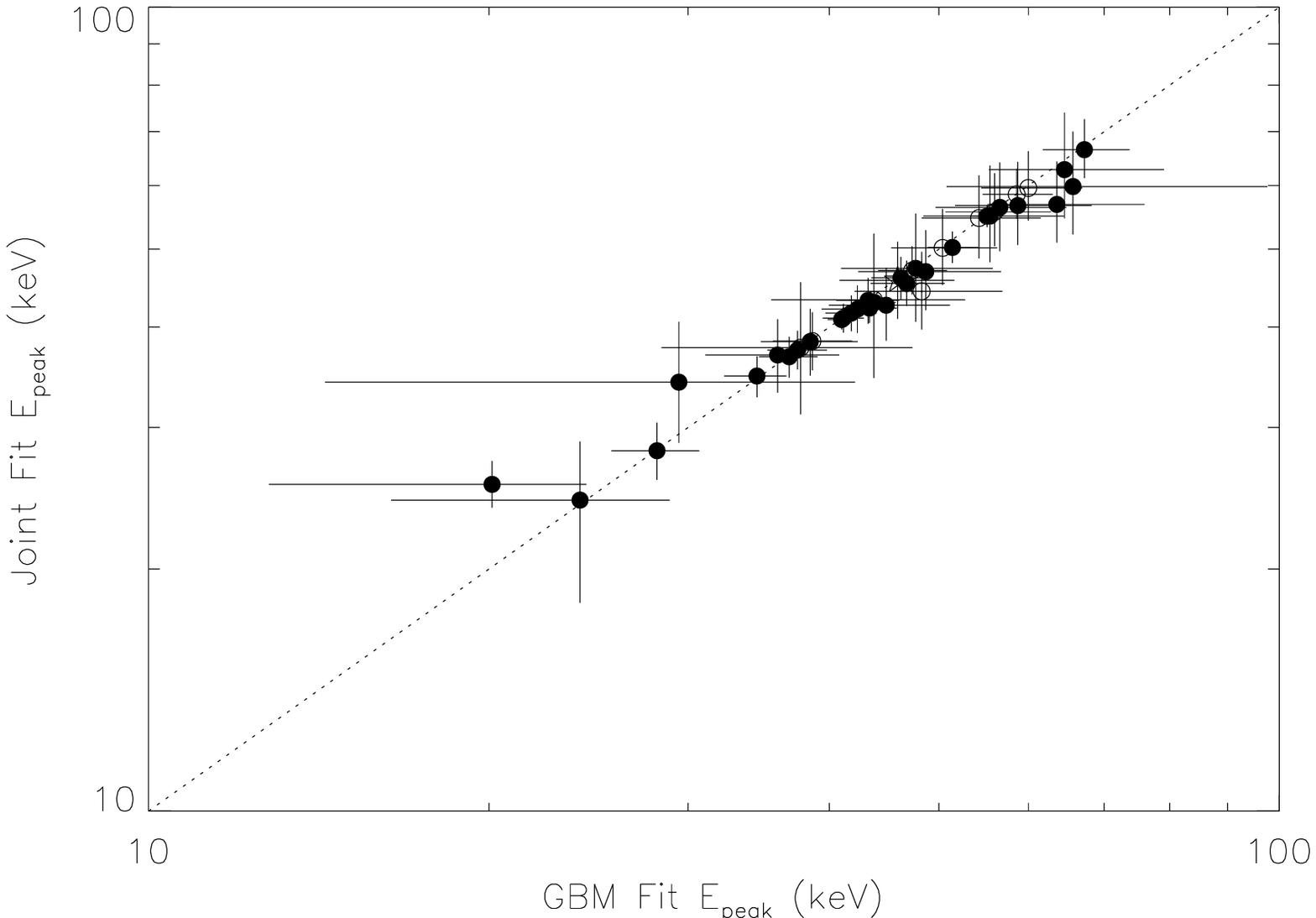}\\
\includegraphics*[scale=0.4]{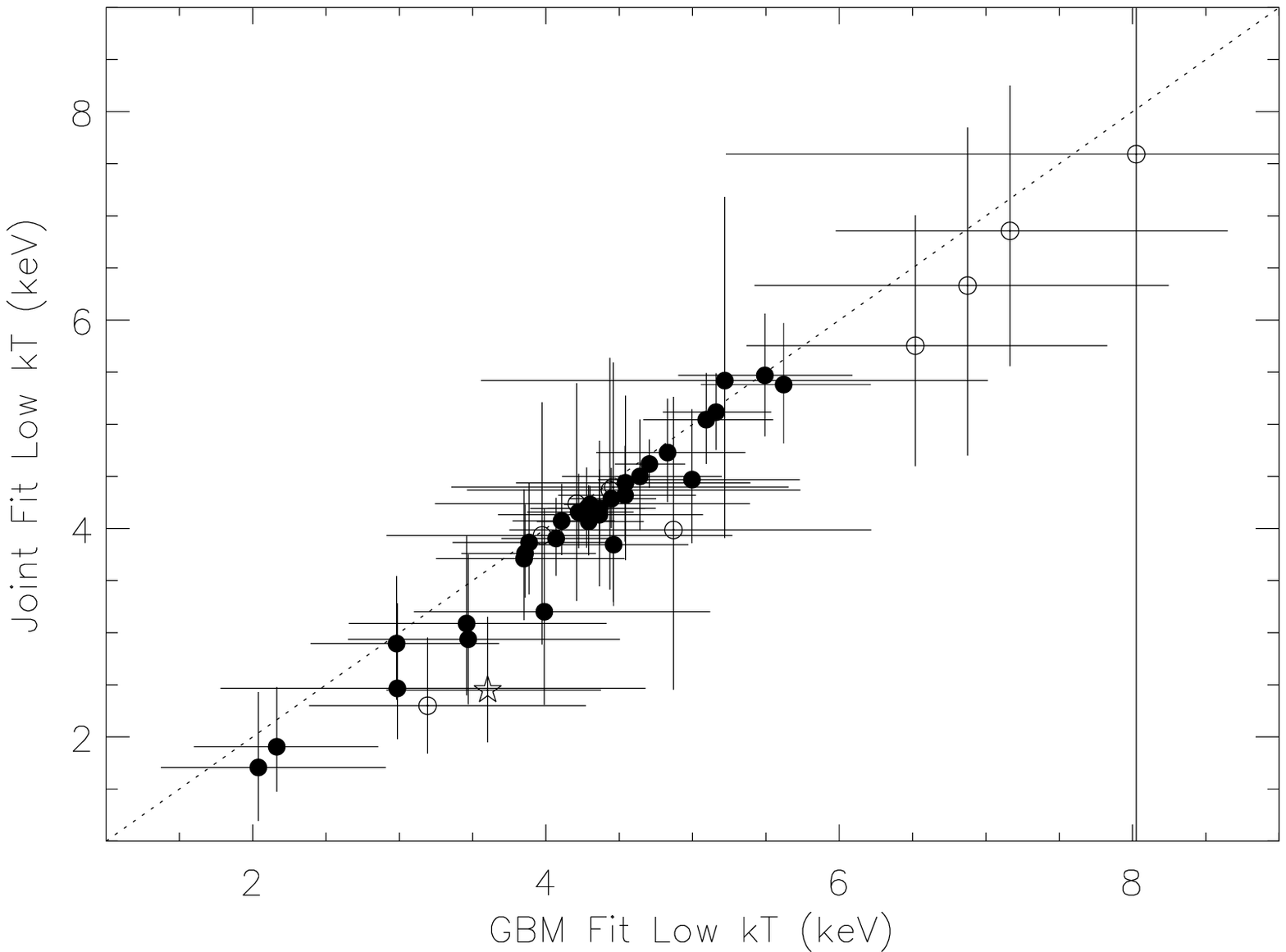}
\includegraphics*[scale=0.4]{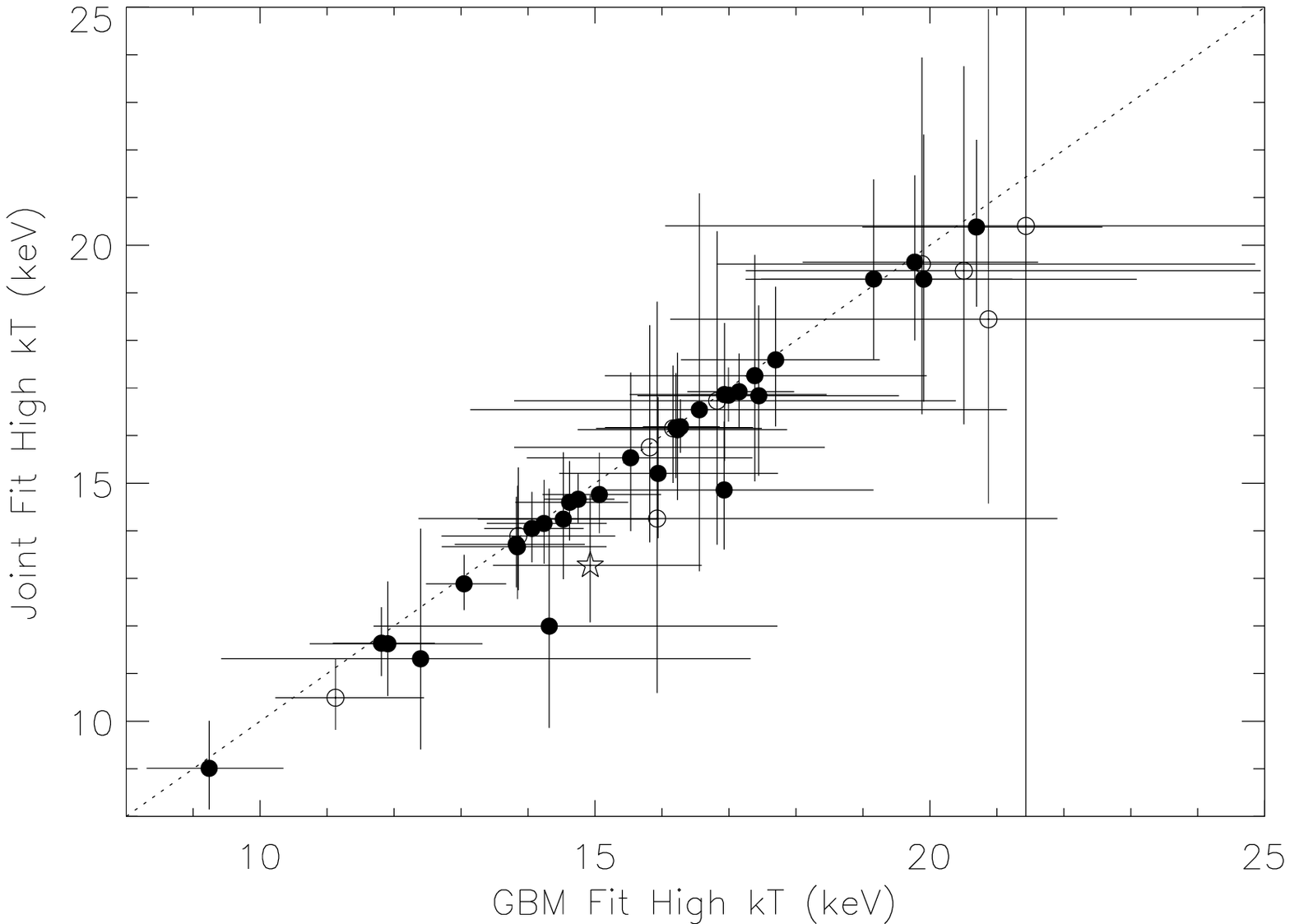}\\
\includegraphics*[scale=0.4]{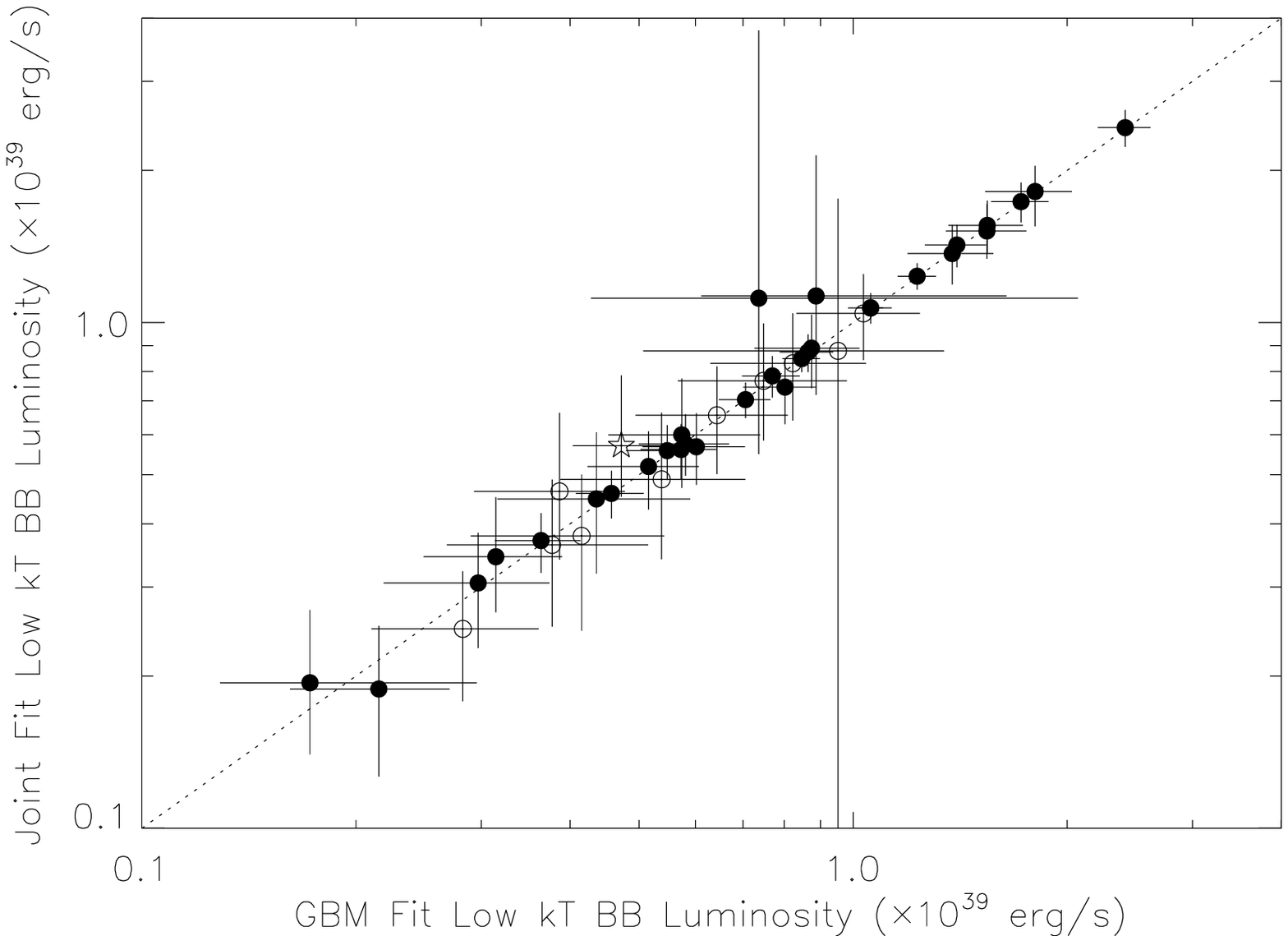}
\includegraphics*[scale=0.4]{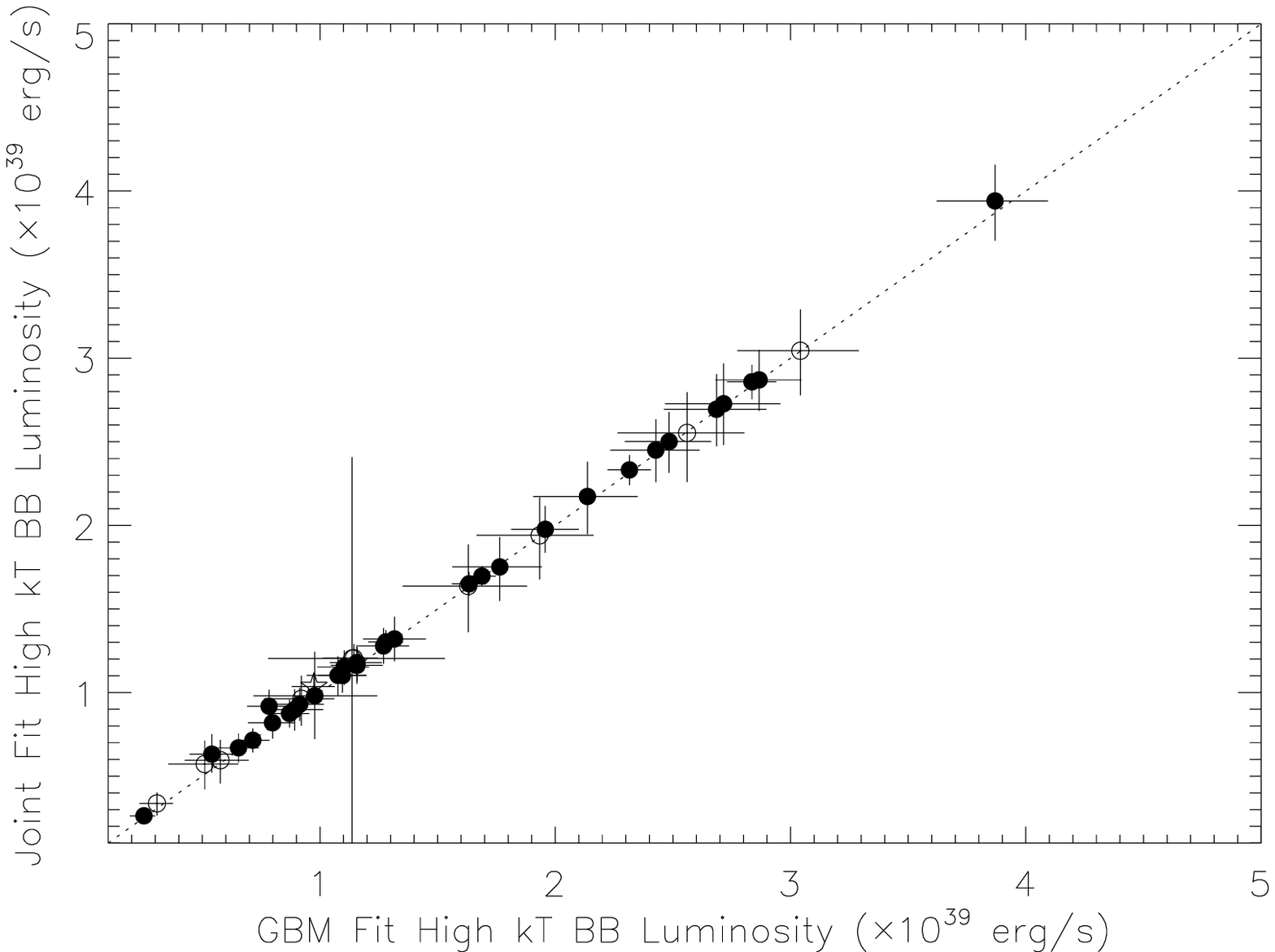}\\
\includegraphics*[scale=0.4]{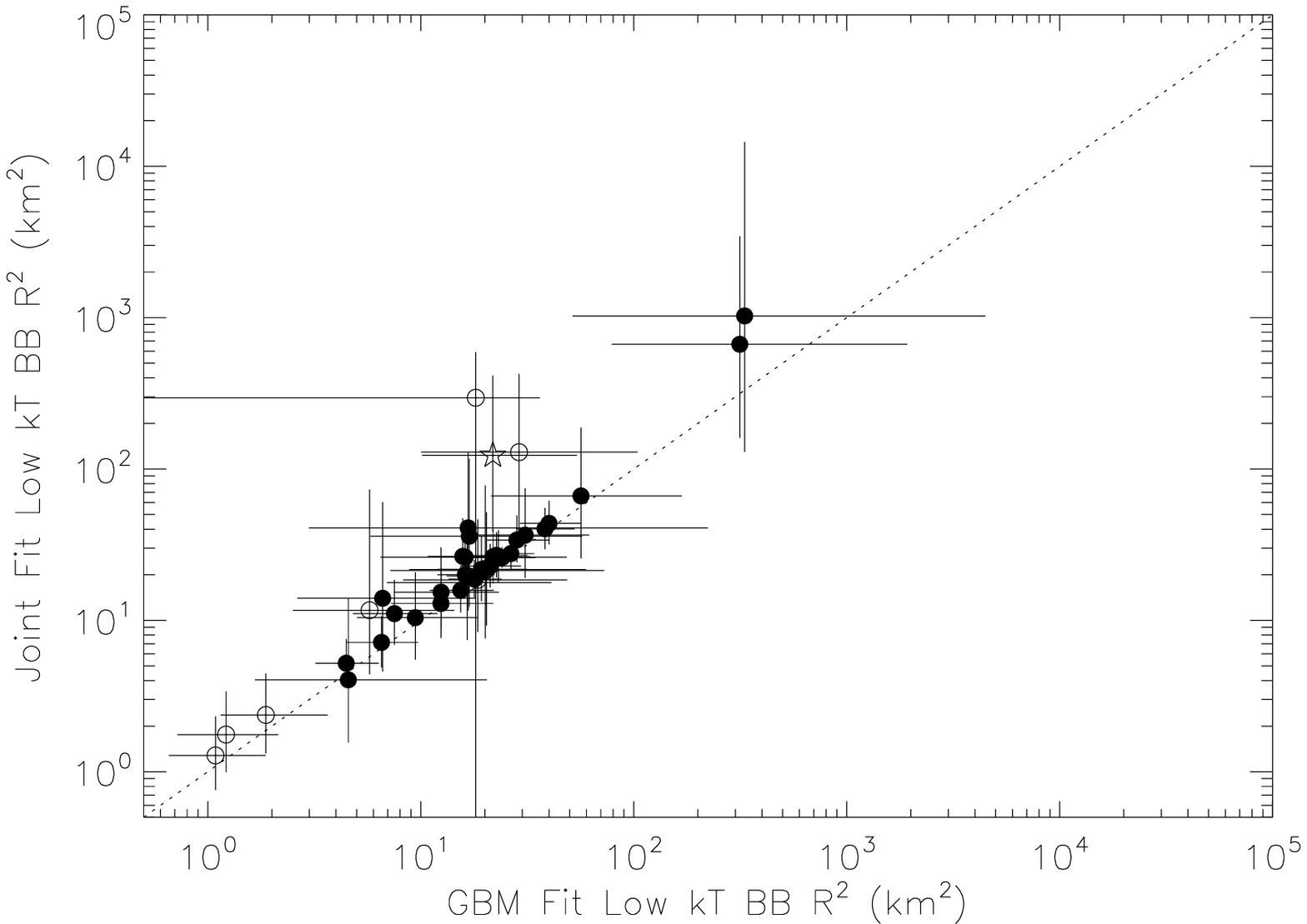}
\includegraphics*[scale=0.4]{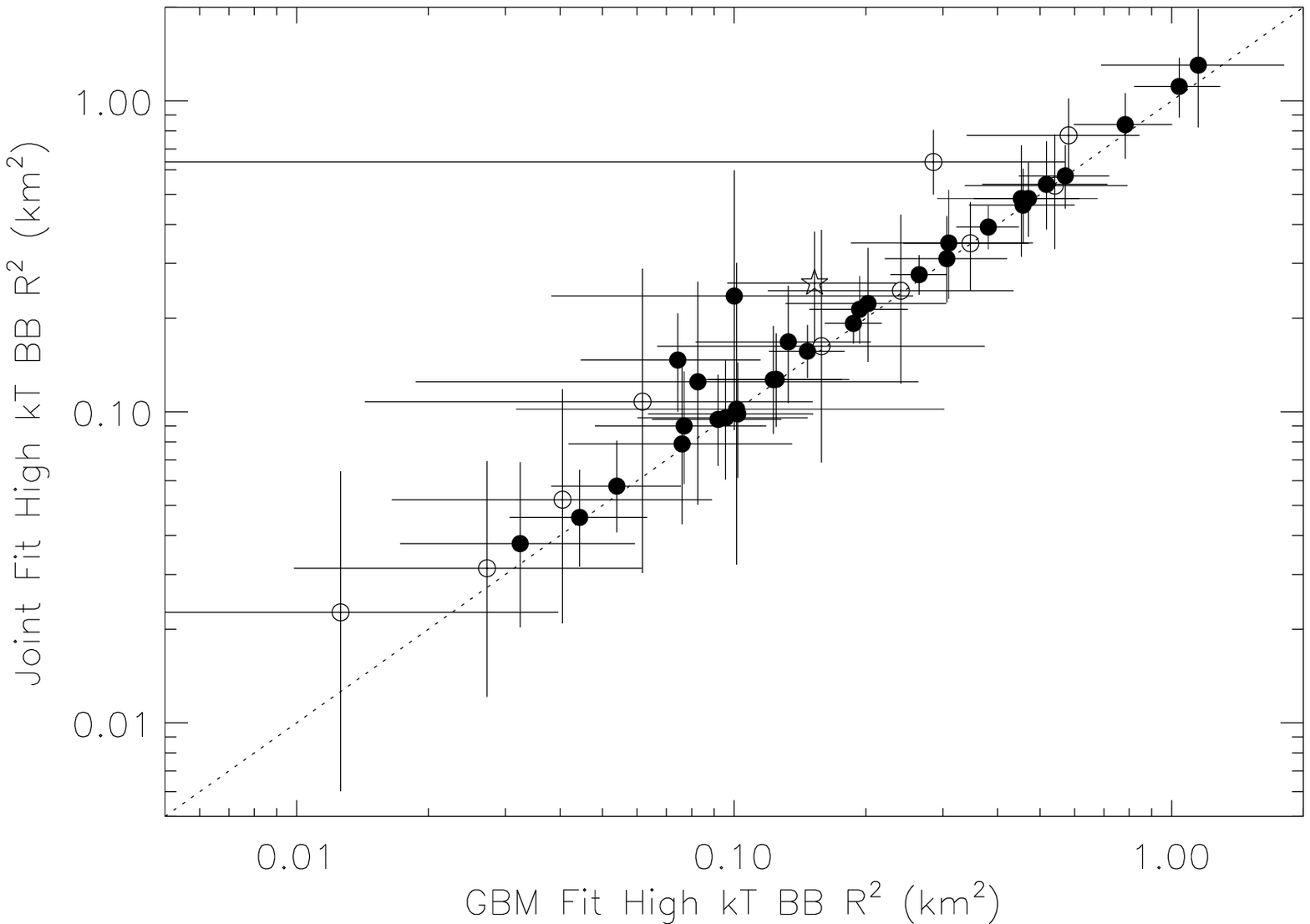}
\caption{Plots of correlations between model parameters obtained by the joint XRT-GBM fits and the parameters obtained by fitting the GBM data only. The black dots are the BB$+$BB bursts, the open circles are the intermediate group bursts and the five point star is the COMPT burst. The dotted lines represent the $x=y$ trend.  \label{fig:gbm-joint}}
\end{figure*}

We have used two different instruments, the \textit{Swift}/XRT and the \textit{Fermi}/GBM, to perform a time-integrated broadband spectral analysis of 42 common events from \sgrnos. By adding the XRT data, we extended the lower energy bound of our earlier spectral analysis \citep{avdh2012} from $\sim 10$\,keV to $\sim 0.5$\,keV. For most model parameters, our joint fit results agree well with the results from fitting the GBM spectra only, as shown in Figure \ref{fig:gbm-joint}. The correlation coefficients between all but one (the COMPT power law index) model parameters derived with and without the XRT data are larger than $0.94$, corresponding to a probability smaller than $3.3\times10^{-10}$. The average COMPT index without the XRT data is $-0.87\pm0.05$, consistent within errors with the mean of $\sim$-0.92 obtained from a much larger burst sample \citep{avdh2012}. However, the inclusion of the XRT data better constrains the COMPT indices, which become harder than the ones derived from spectral fits to the GBM data alone, as shown in Figure \ref{fig:comptpar}. Therefore, we conclude that the COMPT fit to the GBM data only overestimates the emission in the lower energy bands. 

Our analysis provides an important diagnostic for the model preference between the COMPT and BB$+$BB models. By adding the XRT data, we find that 31 of 42 bursts are statistically better described by BB$+$BB. This fact, combined with the observed steepening of COMPT model indices when excluding XRT data, highlight the generic broad curvature of the \sgr burst spectra. We note here that the joint analysis of XRT and GBM spectra is limited to an absorbed energy fluence range of $8.3\times10^{-8}-1.5\times10^{-6}$ erg cm$^{-2}$ ($0.5-200$\,keV), since the brighter events would cause pile-up in the XRT data and the dimmer events would not yield high enough statistics in the XRT data for a constraining spectral analysis. We discuss in the next session the theoretical implications of these results. We also noted the fact that \citet{israel2008} investigated broadband spectral properties of a very rare event (the storm) from SGR\,1900$+$14 while our investigations are about much more common typical short bursts. Therefore, our results extend the \citet{israel2008} results to the more common magnetar outbursts. 

We investigated here \sgr burst spectra in a time-integrated manner. As seen in the bottom panel of Figure \ref{fig:lightcurve}, the hardness ratios of the three parts of the burst, designated by the three peaks in the XRT and GBM lightcurves, show a clear hard to soft spectral evolution. Detailed time-resolved spectral analysis of SGR bursts would provide important insight to the spectral evolution of SGR bursts with time. 

\subsection{The BB$+$BB Model}

The most plausible interpretation of the BB$+$BB model is the emission originating from two hot spots with different temperatures near or on the neutron star surface or in its magnetosphere where local thermodynamic equilibria are achieved. It must be emphasized that a BB spectral fit is an idealization to that emitted by the physical environment of a real photosphere. Due to possible gradients of temperature with optical depth into an evolving region that is approximately in local thermodynamic equilibrium, significant distortions from a true blackbody form are predicted in SGR photospheric spectral models \citep[e.g.][]{ulmer1994,td95,lyubarsky2002,israel2008}.

\begin{figure}
\includegraphics{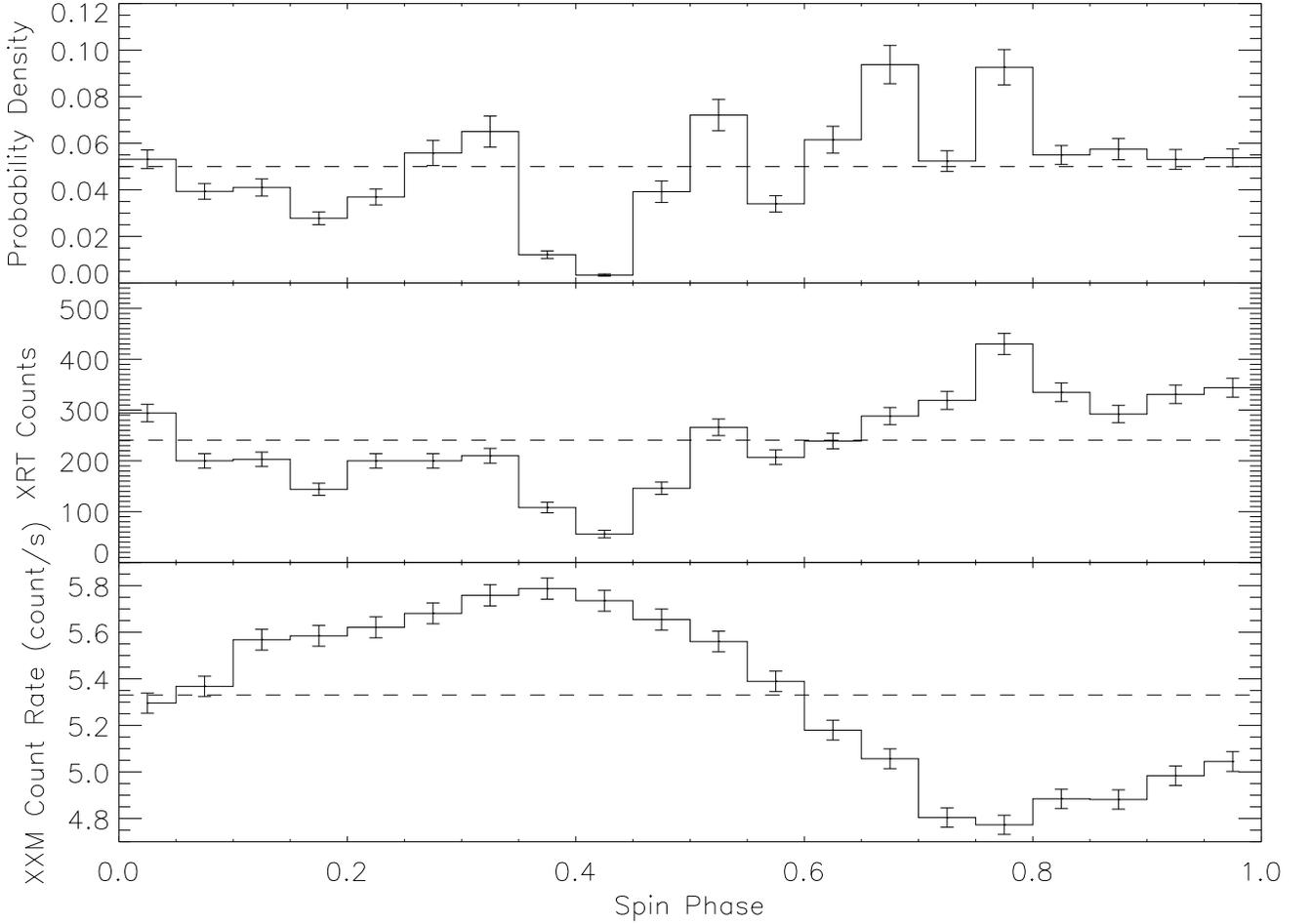}
\caption{\textit{Top}: Phase probability density profile of the BB$+$BB group bursts from \sgr; \textit{middle}: count phase distribution; \textit{bottom}: persistent emission pulse profile from contemporaneous \textit{XMM} observations in the 0.5$-$10 keV band . The horizontal dashed lines in each panel represent the mean value of the burst phase probability density (top panel), the burst phase counts (middle panel) and the persistent emission count rate (bottom panel). \label{fig:phase}}
\end{figure}

To better understand the BB$+$BB behavior and uncover its relation with the spin properties of \sgr, we investigated the phase characteristics of the 31 BB$+$BB bursts, as follows: we first selected all XRT counts collected during 31 burst intervals and converted their arrival times from the \textit{Swift} mission time to the corresponding time at the Solar system barycenter. We then calculated the spin phase for each burst count using the appropriate spin ephemeris of epoch (MJD) 54854 as reported by \citet{dib2012} using both \textit{RXTE} and \textit{Swift} observations. We present the phase distribution of burst counts in the middle panel of Figure \ref{fig:phase}. To ensure that the distribution is not dominated by the excessive counts of the brightest bursts, we also calculated the probability density for each phase bin, which is the average of the normalized (by total counts) phase distributions for all bursts, as shown in the top panel of Figure \ref{fig:phase}. We find that the probability distribution of the burst counts is not uniform over the spin phase of \sgr and the deviation from the mean probability is significant: we calculate the root-mean-square deviation of the phase probability density function from its mean as $0.021 \pm 0.001$. We also compared the phase probability density function to the persistent emission phase profile (bottom panel in Figure \ref{fig:phase}) obtained using contemporaneous \textit{XMM} observations \citep{dib2012}. The phase probability density function is marginally anti-correlated with the persistent emission phase profile in our burst sample with the correlation factor of $-0.5$ corresponding to a chance probability of $3.4 \times 10^{-2}$. This indicates that the burst emission regions on the neutron star surface are not necessarily associated with the site persistently emitting in X-rays (typically a BB with a temperature of 0.5 keV). This is in agreement with the crustal fracturing mechanism for SGR bursts \citep{td95,braithwaite2006,perna2011} as any portion of the solid crust can fracture if the magnetic stress built up is near the threshold to rupture. We also find that the burst probability of some spin phases in \sgr is higher. This could be attributed to a non-uniform surface magnetic field, with some regions having larger magnetic stresses than others.

\subsection{The COMPT model}

A Comptonization spectrum emerges when low energy photons are repeatedly upscattered by the thermal electrons in a corona until the photon energy reaches $E \sim kT_{\rm e}$. We find that the mean value of the Comptonization peak energy, $E_{\rm peak}$ is 44.8 keV, which indicates an average temperature of the thermal electrons, $\langle T_{\rm e} \rangle \sim 5.1\times10^8$\,K. In other words, the average speed of the electrons in the corona is $\sim$ 0.4 c, where c is the speed of light.

The essence of Comptonization spectra is discussed at some length in \citet{lin2011}, in the context of GBM observations of SGR\,J0501$+$4516 bursts. While the turnover energy provides a diagnostic on the hot electron temperature, the power-law slope below the $\nu F_{\nu}$ peak energy defines a measure of the opacity in the Comptonizing region. Specifically, in the simplest theoretical constructs \citep[e.g., see][]{rybicki1979} the index $\lambda$ for a differential photon spectrum $dN/dE \propto E^{\lambda}$ depends only on the magnetic Compton $y$-parameter $y_B=4kT_e/(m_ec^2) \max \{\tau_B ,\, \tau_B^2 \}$ via $\lambda = 1/2 - \sqrt{9/4 + 4/y_B}$. Here $\max\{\tau,\tau^2\}$ is the mean number of scatterings per photon by the hot electrons, where $\tau$ is the effective optical depth for scattering, which in our case is modified by the strong magnetic field and thus dubbed the magnetic optical depth and denoted by $\tau_B$. Since we require the effective value of $\tau_B$, one needs to calculate some Rosseland-type mean opacity, averaged over photon angles and polarizations and accounting for the effects of the strong magnetic fields present, such as anisotropy and polarization-mode switching through scattering in non-uniform $B$ \citep[see also the discussion in][]{lin2011}. This index is realized only in the energy range somewhat above the soft photon injection energy, $E_X$, presumed to be surface thermal X-rays, and somewhat below the characteristic energy associated with the hot thermal electrons, in this case marked via the $\nu F_{\nu}$ peak energy $E_{\rm peak}$. Moreover, the above relationship for $y_B$ requires the scattering to be in the Thomson regime, and the mean photon energy to be lower than that of the electrons.

It is evident that $y_B\gg 1$ cases yield the flattest Comptonized spectra with index around $\lambda \sim -1$. The COMPT fit indices in Table \ref{tab:gbm-xrt} as well as its distribution in the left panel of Figure \ref{fig:comptpar} are nearly always harder than this, indicating that repeated Compton upscattering has difficulty in generating the observed flat spectra.  We note parenthetically, that this was also the case for around $1/3 - 1/2$ the bursts reported for SGR\,J0501$+$4516 in \citet{lin2011}. The inclusion of XRT data in the current work extends the spectral coverage to much lower energies thus enabling us to determine the value of $\lambda$ significantly better than in previous works. This results in systematically higher values, corresponding to a harder spectral slope, with a mean value of $-0.6\pm 0.10$ instead of $-0.92\pm 0.05$, which has made this problem worse. The fact that large $y_B$ is demanded in this fitting protocol would suggest high opacity and strong thermalization might be active in the burst emission region. In such cases, the above dependence of $y_B$ on $T_e$ and $\tau_B$ is not operable, and the Comptonization is saturated, and described instead by a modified Wien or modified BB spectrum \citep[e.g., see][]{ rybicki1979}.  Moreover, the upscattered power-law photon population is generally substantially lower in total number relative to the seed surface X-ray population (i.e., it presents itself as an X-ray tail), a broadband spectral shape that is at odds with the spectral curvature inferred from the fits here.  Hence, there is no strong mandate to prefer a classical, unsaturated Comptonization model for the bursts reported here. Accordingly, the spectra might naturally be expected to exhibit more truly thermalized character, for example a two-blackbody or multi-blackbody signal emanating from a $\tau_B\gg 1$ zone, possibly eliciting spectral distortion imposed by transport within the photosphere \citep[e.g. see][]{ulmer1994,lyubarsky2002}. Perhaps this is what this broad band XRT/GBM analysis has enabled for a magnetar for the first time: the clear discrimination between COMPT and BB$+$BB spectral models.


\acknowledgments 

L.L. is funded through the Post-Doctoral Research Fellowship of the Turkish Academy of Science. MGB acknowledges support from NASA through grant NNX10AC59A. J.G. is supported by the ERC advanced research grant 'GRBs'. C.K. is partially funded by the GBM/Magnetar Key Project (NASA grant NNH07ZDA001-GLAST, PI: C. Kouveliotou). A.L.W. acknowledges support from a Netherlands Organization for Scientific Research (NWO) Vidi Grant.

\clearpage

\begin{deluxetable}{llllllllllllllll}
\tablecolumns{16}
\tabletypesize{\scriptsize}
\setlength{\tabcolsep}{0.025in} 
\rotate
\tablecaption{Spectral fit result of GBM-XRT common bursts from \sgrnos. \label{tab:gbm-xrt}}
\tablewidth{0pt}
\tablehead{
\colhead{ } & \colhead{ } & \colhead{ } & \colhead{COMPT} & \colhead{COMPT} & \colhead{COMPT} & \colhead{cool BB} & \colhead{hot BB} & \colhead{cool BB} & \colhead{hot BB} & \colhead{cool BB} & \colhead{hot BB} & \colhead{BB+BB} & \colhead{ } & \colhead{ } & \colhead{ } \\
\colhead{\#} & \colhead{$T_{\rm start}$\tablenotemark{a}} & \colhead{dt} & \colhead{$\lambda$} & \colhead{$E_{\rm peak}$} & \colhead{$\chi^2/dof$} & \colhead{$kT$} & \colhead{$kT$} & \colhead{$L_{39}$\tablenotemark{b}} & \colhead{$L_{39}$\tablenotemark{b}} & \colhead{$R^2$} & \colhead{$R^2$} & \colhead{$\chi^2/dof$} & \colhead{$F_{\rm GBM}$\tablenotemark{c}} & \colhead{$F_{\rm XRT}$\tablenotemark{d}} & \colhead{$p$} \\
\colhead{} & \colhead{(UTC)} & \colhead{(s)} & \colhead{} & \colhead{(keV)} & \colhead{} & \colhead{(keV)} & \colhead{(keV)} & \colhead{}& \colhead{}& \colhead{(km$^2$)} & \colhead{(km$^2$)} & \colhead{} & \colhead{} & \colhead{} & \colhead{} \\
}
\startdata
01 & 02:34:28.194 & 1.248 & $-0.27^{+ 0.14}_{- 0.14}$ & $42.24^{+ 1.86}_{- 1.78}$ & $191.248/132$ & $ 4.07^{+ 0.35}_{- 0.32}$ & $14.76^{+ 0.88}_{- 0.81}$ & $   0.70^{+   0.06}_{-   0.06}$ & $   1.30^{+   0.07}_{-   0.07}$ & $  20.02^{+   7.40}_{-   5.23}$ & $   0.21^{+   0.06}_{-   0.05}$ &  $158.980/131$ & $   7.51^{+   0.25}_{-   0.28}$ & $   0.92^{+   0.04}_{-   0.07}$ & 0.9996 \\
02 & 02:34:39.794 & 1.152 & $-0.42^{+ 0.41}_{- 0.40}$ & $34.16^{+ 6.45}_{- 5.45}$ & $ 32.945/ 39$ & $ 2.47^{+ 0.82}_{- 0.49}$ & $11.31^{+ 2.73}_{- 1.91}$ & $   0.19^{+   0.08}_{-   0.05}$ & $   0.26^{+   0.04}_{-   0.04}$ & $  40.83^{+  87.84}_{-  29.22}$ & $   0.12^{+   0.14}_{-   0.07}$ & $ 29.642/ 38$ & $   1.28^{+   0.12}_{-   0.18}$ & $   0.22^{+   0.01}_{-   0.09}$ & 0.9766 \\
03 & 02:45:53.041 & 0.232 & $-0.33^{+ 0.19}_{- 0.18}$ & $41.04^{+ 1.71}_{- 1.68}$ & $ 81.479/ 68$ & $ 5.04^{+ 0.45}_{- 0.42}$ & $14.60^{+ 0.86}_{- 0.80}$ & $   1.56^{+   0.19}_{-   0.18}$ & $   2.69^{+   0.21}_{-   0.22}$ & $  18.71^{+   6.39}_{-   4.57}$ & $   0.46^{+   0.14}_{-   0.11}$ & $ 69.707/ 67$ & $   3.06^{+   0.10}_{-   0.12}$ & $   0.19^{+   0.02}_{-   0.03}$ & 0.9305 \\
04 & 02:53:45.849 & 1.008 & $-0.21^{+ 0.29}_{- 0.28}$ & $44.31^{+ 5.30}_{- 4.59}$ & $ 43.958/ 50$ & $ 5.75^{+ 1.25}_{- 1.15}$ & $18.44^{+ 6.52}_{- 3.87}$ & $   0.25^{+   0.07}_{-   0.07}$ & $   0.34^{+   0.06}_{-   0.07}$ & $   1.76^{+   1.64}_{-   0.77}$ & $   0.02^{+   0.04}_{-   0.02}$ & $ 46.837/ 49$ & $   1.85^{+   0.15}_{-   0.17}$ & $   0.50^{+   0.03}_{-   0.06}$ & 0.6609 \\
05 & 02:55:15.993 & 0.360 & $-0.29^{+ 0.37}_{- 0.34}$ & $54.66^{+ 7.11}_{- 5.95}$ & $ 24.680/ 28$ & $ 6.86^{+ 1.39}_{- 1.30}$ & $19.60^{+ 4.34}_{- 3.15}$ & $   0.36^{+   0.13}_{-   0.11}$ & $   0.60^{+   0.12}_{-   0.14}$ & $   1.28^{+   1.04}_{-   0.52}$ & $   0.03^{+   0.04}_{-   0.02}$ & $ 26.868/ 27$ & $   1.11^{+   0.11}_{-   0.11}$ & $   0.30^{+   0.03}_{-   0.06}$ & 0.5153 \\
06 & 02:56:52.649 & 0.320 & $-0.32^{+ 0.38}_{- 0.36}$ & $38.36^{+ 3.77}_{- 3.55}$ & $ 42.710/ 32$ & $ 3.09^{+ 0.84}_{- 0.69}$ & $11.63^{+ 1.31}_{- 1.10}$ & $   0.31^{+   0.08}_{-   0.08}$ & $   0.82^{+   0.09}_{-   0.10}$ & $  26.13^{+   9.00}_{-   6.13}$ & $   0.35^{+   0.17}_{-   0.12}$ & $ 41.331/ 31$ & $   1.04^{+   0.07}_{-   0.10}$ & $   0.32^{+   0.02}_{-   0.11}$ & 0.9424 \\
07 & 02:56:53.705 & 0.288 & $-0.79^{+ 0.35}_{- 0.33}$ & $62.79^{+11.14}_{- 8.19}$ & $ 24.128/ 29$ & $ 2.94^{+ 0.81}_{- 0.62}$ & $16.84^{+ 1.90}_{- 1.68}$ & $   0.34^{+   0.11}_{-   0.08}$ & $   0.93^{+   0.10}_{-   0.10}$ & $  35.96^{+  80.44}_{-  23.69}$ & $   0.09^{+   0.04}_{-   0.03}$ & $ 22.212/ 28$ & $   1.06^{+   0.08}_{-   0.11}$ & $   0.20^{+   0.01}_{-   0.07}$ & 0.9814 \\
08 & 02:57:18.393 & 0.088 & $-1.35^{+ 0.39}_{- 0.34}$ & $43.26^{+ 9.02}_{- 8.69}$ & $ 10.233/ 19$ & $ 4.40^{+ 1.19}_{- 1.10}$ & $16.73^{+ 3.56}_{- 3.02}$ & $   1.04^{+   0.20}_{-   0.20}$ & $   1.64^{+   0.25}_{-   0.28}$ & $  21.68^{+  29.97}_{-  12.43}$ & $   0.16^{+   0.22}_{-   0.09}$ & $ 12.402/ 18$ & $   0.71^{+   0.06}_{-   0.07}$ & $   0.13^{+   0.03}_{-   0.04}$ & 0.3987 \\
09 & 04:08:31.630 & 0.208 & $-0.44^{+ 0.28}_{- 0.26}$ & $46.06^{+ 2.85}_{- 2.76}$ & $ 60.365/ 51$ & $ 3.71^{+ 0.67}_{- 0.59}$ & $14.15^{+ 0.91}_{- 0.84}$ & $   0.89^{+   0.15}_{-   0.15}$ & $   2.50^{+   0.18}_{-   0.19}$ & $  36.54^{+  38.09}_{-  17.46}$ & $   0.49^{+   0.15}_{-   0.12}$ & $ 55.320/ 50$ & $   2.14^{+   0.11}_{-   0.12}$ & $   0.22^{+   0.02}_{-   0.05}$ & 0.9465 \\
10 & 04:10:59.166 & 0.208 & $-0.53^{+ 0.38}_{- 0.34}$ & $55.63^{+ 6.50}_{- 5.19}$ & $ 22.919/ 34$ & $ 7.59^{+ 7.59}_{- 7.59}$ & $20.40^{+20.40}_{-20.40}$ & $   0.88^{+   0.88}_{-   0.88}$ & $   1.20^{+   1.20}_{-   1.20}$ & $ 295.40^{+ 295.40}_{- 295.40}$ & $   0.64^{+   0.17}_{-   0.14}$ & $ 26.889/ 33$ & $   1.40^{+   0.10}_{-   0.11}$ & $   0.22^{+   0.03}_{-   0.04}$ & 0.6539 \\
11 & 04:19:27.593 & 0.264 & $-0.99^{+ 0.18}_{- 0.17}$ & $34.77^{+ 2.01}_{- 2.06}$ & $ 81.608/ 76$ & $ 4.19^{+ 0.39}_{- 0.38}$ & $13.72^{+ 1.00}_{- 0.90}$ & $   1.73^{+   0.16}_{-   0.16}$ & $   2.45^{+   0.18}_{-   0.19}$ & $  43.69^{+  17.94}_{-  11.93}$ & $   0.54^{+   0.20}_{-   0.15}$ & $ 80.685/ 75$ & $   3.27^{+   0.11}_{-   0.12}$ & $   0.42^{+   0.04}_{-   0.05}$ & 0.9831 \\
12 & 04:21:32.313 & 0.312 & $-0.21^{+ 0.14}_{- 0.14}$ & $40.87^{+ 1.17}_{- 1.15}$ & $115.983/101$ & $ 4.32^{+ 0.47}_{- 0.44}$ & $12.89^{+ 0.61}_{- 0.55}$ & $   1.52^{+   0.20}_{-   0.18}$ & $   3.94^{+   0.22}_{-   0.24}$ & $  33.95^{+  15.31}_{-   9.77}$ & $   1.11^{+   0.26}_{-   0.23}$ & $114.304/100$ & $   5.24^{+   0.13}_{-   0.14}$ & $   0.44^{+   0.04}_{-   0.05}$ & 0.9735 \\
13 & 04:21:41.825 & 0.192 & $-0.83^{+ 0.39}_{- 0.36}$ & $56.36^{+ 7.73}_{- 6.63}$ & $ 38.622/ 27$ & $ 2.90^{+ 0.65}_{- 0.54}$ & $16.86^{+ 1.50}_{- 1.40}$ & $   0.60^{+   0.18}_{-   0.13}$ & $   1.32^{+   0.13}_{-   0.13}$ & $  66.36^{+ 121.03}_{-  40.68}$ & $   0.13^{+   0.05}_{-   0.04}$ & $ 29.603/ 26$ & $   1.04^{+   0.08}_{-   0.10}$ & $   0.20^{+   0.02}_{-   0.07}$ & 0.9855 \\
14 & 04:21:49.321 & 0.352 & $-0.68^{+ 0.33}_{- 0.31}$ & $38.42^{+ 3.31}_{- 3.07}$ & $ 41.476/ 44$ & $ 2.30^{+ 0.65}_{- 0.46}$ & $10.49^{+ 0.82}_{- 0.68}$ & $   0.46^{+   0.20}_{-   0.12}$ & $   1.21^{+   0.08}_{-   0.09}$ & $ 129.11^{+ 296.00}_{-  88.02}$ & $   0.77^{+   0.24}_{-   0.22}$ & $ 43.926/ 43$ & $   1.57^{+   0.07}_{-   0.12}$ & $   0.27^{+   0.01}_{-   0.10}$ & 0.7148 \\
15 & 04:23:01.345 & 0.168 & $-0.91^{+ 0.18}_{- 0.17}$ & $36.74^{+ 2.15}_{- 2.15}$ & $ 75.996/ 62$ & $ 5.12^{+ 0.37}_{- 0.36}$ & $16.17^{+ 1.14}_{- 1.06}$ & $   2.43^{+   0.20}_{-   0.20}$ & $   2.73^{+   0.24}_{-   0.25}$ & $  27.60^{+   8.05}_{-   5.97}$ & $   0.31^{+   0.12}_{-   0.09}$ & $ 66.447/ 61$ & $   2.65^{+   0.09}_{-   0.11}$ & $   0.23^{+   0.04}_{-   0.04}$ & 0.9622 \\
16 & 04:23:35.961 & 0.112 & $-0.79^{+ 0.37}_{- 0.33}$ & $50.17^{+ 5.94}_{- 5.04}$ & $ 26.312/ 24$ & $ 4.37^{+ 1.27}_{- 0.95}$ & $15.75^{+ 2.57}_{- 1.99}$ & $   0.83^{+   0.21}_{-   0.19}$ & $   1.94^{+   0.23}_{-   0.26}$ & $  17.74^{+  26.40}_{-  10.33}$ & $   0.25^{+   0.19}_{-   0.12}$ & $ 26.934/ 23$ & $   0.96^{+   0.06}_{-   0.08}$ & $   0.14^{+   0.03}_{-   0.05}$ & 0.4023 \\
17 & 04:23:56.473 & 0.240 & $-1.09^{+ 0.37}_{- 0.33}$ & $47.33^{+ 8.03}_{- 6.71}$ & $ 35.994/ 28$ & $ 4.44^{+ 0.84}_{- 0.74}$ & $17.26^{+ 2.53}_{- 2.22}$ & $   0.52^{+   0.09}_{-   0.09}$ & $   0.90^{+   0.12}_{-   0.13}$ & $  10.42^{+  10.35}_{-   4.92}$ & $   0.08^{+   0.06}_{-   0.04}$ & $ 35.231/ 27$ & $   1.03^{+   0.08}_{-   0.09}$ & $   0.19^{+   0.03}_{-   0.05}$ & 0.9042 \\
18 & 04:23:58.913 & 0.408 & $-1.05^{+ 0.25}_{- 0.24}$ & $45.74^{+ 5.34}_{- 4.80}$ & $ 49.176/ 46$ & $ 2.45^{+ 0.70}_{- 0.50}$ & $13.27^{+ 1.47}_{- 1.19}$ & $   0.57^{+   0.21}_{-   0.12}$ & $   1.04^{+   0.08}_{-   0.08}$ & $ 122.96^{+ 290.05}_{-  84.30}$ & $   0.26^{+   0.12}_{-   0.09}$ & $ 53.494/ 45$ & $   1.69^{+   0.09}_{-   0.13}$ & $   0.30^{+   0.02}_{-   0.08}$ & 0.9084 \\
19 & 04:29:51.974 & 0.104 & $-0.15^{+ 0.36}_{- 0.33}$ & $58.50^{+ 4.47}_{- 3.94}$ & $ 26.192/ 27$ & $ 3.93^{+ 1.28}_{- 1.05}$ & $16.15^{+ 1.32}_{- 1.14}$ & $   0.66^{+   0.16}_{-   0.15}$ & $   3.04^{+   0.25}_{-   0.27}$ & $  21.33^{+  56.29}_{-  13.73}$ & $   0.35^{+   0.12}_{-   0.10}$ & $ 23.320/ 26$ & $   1.21^{+   0.07}_{-   0.09}$ & $   0.12^{+   0.03}_{-   0.05}$ & 0.8910 \\
20 & 04:32:04.678 & 0.608 & $-0.85^{+ 0.32}_{- 0.30}$ & $46.88^{+ 5.92}_{- 4.91}$ & $ 39.847/ 46$ & $ 3.87^{+ 0.57}_{- 0.50}$ & $15.53^{+ 1.79}_{- 1.54}$ & $   0.37^{+   0.05}_{-   0.05}$ & $   0.71^{+   0.07}_{-   0.07}$ & $  12.92^{+   9.36}_{-   5.29}$ & $   0.10^{+   0.05}_{-   0.04}$ & $ 29.538/ 45$ & $   1.96^{+   0.12}_{-   0.15}$ & $   0.22^{+   0.02}_{-   0.04}$ & 0.9570 \\
21 & 04:32:10.070 & 0.312 & $-1.14^{+ 0.27}_{- 0.25}$ & $54.97^{+ 8.56}_{- 6.81}$ & $ 51.362/ 37$ & $ 3.76^{+ 0.47}_{- 0.42}$ & $17.59^{+ 1.54}_{- 1.40}$ & $   0.56^{+   0.07}_{-   0.07}$ & $   1.16^{+   0.11}_{-   0.11}$ & $  21.72^{+  14.15}_{-   8.32}$ & $   0.09^{+   0.04}_{-   0.03}$ & $ 42.803/ 36$ & $   1.60^{+   0.10}_{-   0.11}$ & $   0.20^{+   0.03}_{-   0.04}$ & 0.9705 \\
22 & 04:32:16.350 & 0.120 & $-0.39^{+ 0.61}_{- 0.51}$ & $28.07^{+ 2.35}_{- 2.24}$ & $ 42.995/ 28$ & $ 5.47^{+ 0.59}_{- 0.58}$ & $16.54^{+ 4.54}_{- 3.39}$ & $   1.82^{+   0.23}_{-   0.27}$ & $   0.98^{+   0.26}_{-   0.26}$ & $  15.80^{+   6.86}_{-   4.53}$ & $   0.10^{+   0.20}_{-   0.07}$ & $ 37.277/ 27$ & $   1.01^{+   0.07}_{-   0.07}$ & $   0.17^{+   0.03}_{-   0.03}$ & 0.9163 \\
23 & 04:32:18.846 & 1.016 & $-0.89^{+ 0.12}_{- 0.11}$ & $50.25^{+ 2.33}_{- 2.19}$ & $150.368/131$ & $ 4.28^{+ 0.30}_{- 0.28}$ & $16.92^{+ 0.80}_{- 0.75}$ & $   0.85^{+   0.05}_{-   0.05}$ & $   1.65^{+   0.07}_{-   0.07}$ & $  19.59^{+   5.72}_{-   4.30}$ & $   0.16^{+   0.03}_{-   0.03}$ & $134.705/130$ & $   7.73^{+   0.20}_{-   0.21}$ & $   0.88^{+   0.06}_{-   0.07}$ & 0.9999 \\
24 & 04:33:01.510 & 0.232 & $-0.61^{+ 0.22}_{- 0.21}$ & $37.45^{+ 2.12}_{- 2.02}$ & $ 65.511/ 60$ & $ 4.50^{+ 0.55}_{- 0.52}$ & $13.66^{+ 1.28}_{- 1.10}$ & $   1.37^{+   0.19}_{-   0.18}$ & $   2.17^{+   0.21}_{-   0.22}$ & $  25.99^{+  13.37}_{-   8.12}$ & $   0.48^{+   0.23}_{-   0.17}$ & $ 62.295/ 59$ & $   2.49^{+   0.11}_{-   0.11}$ & $   0.34^{+   0.03}_{-   0.05}$ & 0.9675 \\
25 & 04:33:17.486 & 0.576 & $-1.00^{+ 0.23}_{- 0.22}$ & $56.65^{+ 7.56}_{- 6.05}$ & $ 56.285/ 52$ & $ 4.73^{+ 0.52}_{- 0.48}$ & $19.64^{+ 1.82}_{- 1.64}$ & $   0.46^{+   0.05}_{-   0.05}$ & $   0.88^{+   0.08}_{-   0.08}$ & $   7.15^{+   3.49}_{-   2.27}$ & $   0.05^{+   0.02}_{-   0.01}$ & $ 47.499/ 51$ & $   2.37^{+   0.15}_{-   0.17}$ & $   0.35^{+   0.03}_{-   0.05}$ & 0.9824 \\
26 & 04:33:29.134 & 0.168 & $ 0.01^{+ 0.42}_{- 0.38}$ & $43.17^{+ 2.87}_{- 2.78}$ & $ 42.645/ 31$ & $ 1.71^{+ 0.72}_{- 0.51}$ & $11.64^{+ 0.75}_{- 0.69}$ & $   1.12^{+   2.66}_{-   0.57}$ & $   1.98^{+   0.14}_{-   0.14}$ & $1025.33^{+13393.7}_{- 895.54}$ & $   0.84^{+   0.22}_{-   0.19}$ & $ 34.018/ 30$ & $   1.18^{+   0.05}_{-   0.11}$ & $   0.17^{+   0.00}_{-   0.16}$ & 0.9595 \\
27 & 04:33:35.750 & 0.480 & $-0.66^{+ 0.23}_{- 0.22}$ & $42.11^{+ 2.95}_{- 2.77}$ & $ 68.651/ 59$ & $ 4.13^{+ 0.71}_{- 0.69}$ & $14.24^{+ 1.40}_{- 1.26}$ & $   0.58^{+   0.08}_{-   0.08}$ & $   1.18^{+   0.10}_{-   0.11}$ & $  15.35^{+  15.02}_{-   6.50}$ & $   0.22^{+   0.11}_{-   0.08}$ & $ 67.225/ 58$ & $   2.55^{+   0.12}_{-   0.15}$ & $   0.36^{+   0.03}_{-   0.05}$ & 0.9665 \\
28 & 04:33:50.294 & 0.560 & $-0.74^{+ 0.28}_{- 0.26}$ & $42.58^{+ 4.82}_{- 4.11}$ & $ 44.346/ 51$ & $ 5.38^{+ 0.59}_{- 0.56}$ & $19.29^{+ 3.04}_{- 2.58}$ & $   0.56^{+   0.07}_{-   0.07}$ & $   0.67^{+   0.09}_{-   0.09}$ & $   5.21^{+   2.32}_{-   1.52}$ & $   0.04^{+   0.03}_{-   0.02}$ & $ 39.108/ 50$ & $   2.13^{+   0.14}_{-   0.15}$ & $   0.49^{+   0.04}_{-   0.05}$ & 0.9906 \\
29 & 04:34:12.430 & 0.128 & $-0.23^{+ 0.35}_{- 0.32}$ & $47.03^{+ 3.38}_{- 3.10}$ & $ 36.254/ 32$ & $ 4.24^{+ 1.15}_{- 0.93}$ & $13.89^{+ 1.44}_{- 1.14}$ & $   0.77^{+   0.23}_{-   0.18}$ & $   2.55^{+   0.24}_{-   0.29}$ & $  18.51^{+  27.89}_{-  10.12}$ & $   0.53^{+   0.24}_{-   0.20}$ & $ 34.438/ 31$ & $   1.32^{+   0.07}_{-   0.09}$ & $   0.18^{+   0.02}_{-   0.05}$ & 0.8805 \\
30 & 04:34:14.502 & 0.392 & $-0.24^{+ 0.47}_{- 0.42}$ & $56.84^{+7.41}_{- 5.86}$ & $ 28.03/ 22$ & $ 3.20^{+ 0.99}_{- 0.89}$ & $14.86^{+ 1.44}_{- 1.25}$ & $   0.19^{+   0.07}_{-   0.06}$ & $   0.92^{+   0.10}_{-   0.10}$ & $   13.98^{+  46.46}_{-   9.40}$ & $   0.15^{+   0.06}_{-   0.05}$ & $ 25.48/ 21$ & $   3.50^{+   0.28}_{-   0.37}$ & $   0.61^{+   0.04}_{-   0.34}$ & 0.9213 \\
31 & 04:34:20.622 & 1.184 & $-0.36^{+ 0.10}_{- 0.09}$ & $42.43^{+ 1.06}_{- 1.03}$ & $219.531/158$ & $ 4.62^{+ 0.24}_{- 0.23}$ & $14.66^{+ 0.53}_{- 0.50}$ & $   1.23^{+   0.08}_{-   0.07}$ & $   2.33^{+   0.09}_{-   0.09}$ & $  21.13^{+   3.94}_{-   3.25}$ & $   0.39^{+   0.07}_{-   0.06}$ & $165.813/157$ & $  12.98^{+   0.26}_{-   0.26}$ & $   1.42^{+   0.07}_{-   0.08}$ & 0.9998 \\
32 & 04:39:23.636 & 0.432 & $-0.64^{+ 0.25}_{- 0.23}$ & $59.57^{+ 6.60}_{- 5.29}$ & $ 41.121/ 50$ & $ 6.33^{+ 1.52}_{- 1.63}$ & $19.46^{+ 4.29}_{- 3.22}$ & $   0.49^{+   0.17}_{-   0.15}$ & $   0.96^{+   0.14}_{-   0.16}$ & $   2.37^{+   2.08}_{-   1.05}$ & $   0.05^{+   0.07}_{-   0.03}$ & $ 49.488/ 49$ & $   2.01^{+   0.12}_{-   0.14}$ & $   0.35^{+   0.04}_{-   0.05}$ & 0.8472 \\
33 & 04:39:29.908 & 0.184 & $-0.71^{+ 0.19}_{- 0.18}$ & $41.63^{+ 2.16}_{- 2.10}$ & $ 77.969/ 67$ & $ 4.07^{+ 0.35}_{- 0.33}$ & $14.05^{+ 0.77}_{- 0.71}$ & $   1.42^{+   0.14}_{-   0.14}$ & $   2.87^{+   0.18}_{-   0.19}$ & $  40.24^{+  15.06}_{-  10.66}$ & $   0.57^{+   0.15}_{-   0.12}$ & $ 64.717/ 66$ & $   2.38^{+   0.08}_{-   0.10}$ & $   0.31^{+   0.04}_{-   0.04}$ & 0.9585 \\
34 & 04:39:36.628 & 0.336 & $-1.18^{+ 0.23}_{- 0.21}$ & $36.93^{+ 3.98}_{- 3.76}$ & $ 50.047/ 53$ & $ 4.16^{+ 0.37}_{- 0.34}$ & $16.12^{+ 1.61}_{- 1.47}$ & $   0.87^{+   0.07}_{-   0.08}$ & $   1.10^{+   0.10}_{-   0.10}$ & $  22.80^{+   9.08}_{-   6.30}$ & $   0.13^{+   0.06}_{-   0.04}$ & $ 35.201/ 52$ & $   1.95^{+   0.10}_{-   0.12}$ & $   0.35^{+   0.04}_{-   0.05}$ & 0.9943 \\
35 & 05:49:17.289 & 0.536 & $-0.32^{+ 0.20}_{- 0.20}$ & $45.33^{+ 3.05}_{- 2.82}$ & $ 91.248/ 76$ & $ 4.47^{+ 0.68}_{- 0.61}$ & $15.21^{+ 1.61}_{- 1.36}$ & $   0.57^{+   0.09}_{-   0.09}$ & $   1.15^{+   0.10}_{-   0.11}$ & $  11.08^{+   7.40}_{-   4.18}$ & $   0.17^{+   0.09}_{-   0.06}$ & $ 90.915/ 75$ & $   2.83^{+   0.15}_{-   0.15}$ & $   0.49^{+   0.04}_{-   0.05}$ & 0.9868 \\
36 & 05:49:46.105 & 0.200 & $-1.08^{+ 0.49}_{- 0.44}$ & $24.37^{+ 4.45}_{- 6.21}$ & $ 41.745/ 27$ & $ 1.91^{+ 0.57}_{- 0.43}$ & $ 9.01^{+ 1.00}_{- 0.86}$ & $   1.13^{+   1.01}_{-   0.41}$ & $   1.10^{+   0.11}_{-   0.12}$ & $ 666.35^{+2782.43}_{- 505.86}$ & $   1.30^{+   0.67}_{-   0.48}$ & $ 38.411/ 26$ & $   0.87^{+   0.05}_{-   0.12}$ & $   0.11^{+   0.00}_{-   0.09}$ & 0.9913 \\
37 & 06:06:41.596 & 1.000 & $ 0.17^{+ 0.19}_{- 0.19}$ & $25.49^{+ 1.76}_{- 1.65}$ & $ 82.334/ 69$ & $ 3.84^{+ 0.57}_{- 0.59}$ & $12.00^{+ 2.89}_{- 2.14}$ & $   0.75^{+   0.10}_{-   0.12}$ & $   0.63^{+   0.12}_{-   0.11}$ & $  26.46^{+  20.71}_{-   9.90}$ & $   0.24^{+   0.36}_{-   0.15}$ & $ 70.145/ 68$ & $   3.79^{+   0.18}_{-   0.22}$ & $   0.87^{+   0.04}_{-   0.07}$ & 0.9990 \\
38 & 07:26:28.960 & 1.672 & $-0.56^{+ 0.10}_{- 0.10}$ & $42.91^{+ 1.36}_{- 1.31}$ & $317.180/180$ & $ 4.23^{+ 0.18}_{- 0.17}$ & $16.19^{+ 0.58}_{- 0.55}$ & $   1.07^{+   0.04}_{-   0.04}$ & $   1.70^{+   0.06}_{-   0.06}$ & $  25.99^{+   4.42}_{-   3.72}$ & $   0.19^{+   0.03}_{-   0.03}$ & $180.111/179$ & $  13.85^{+   0.28}_{-   0.32}$ & $   1.88^{+   0.08}_{-   0.08}$ & 1.0000 \\
39 & 10:49:29.944 & 0.456 & $-1.25^{+ 0.22}_{- 0.21}$ & $59.82^{+10.20}_{- 7.68}$ & $ 53.961/ 50$ & $ 3.90^{+ 0.39}_{- 0.36}$ & $20.38^{+ 1.83}_{- 1.67}$ & $   0.78^{+   0.07}_{-   0.07}$ & $   1.28^{+   0.11}_{-   0.11}$ & $  26.27^{+  13.39}_{-   8.64}$ & $   0.06^{+   0.02}_{-   0.02}$ & $ 37.806/ 49$ & $   2.76^{+   0.16}_{-   0.16}$ & $   0.31^{+   0.03}_{-   0.04}$ & 0.9998 \\
40 & 10:52:11.888 & 0.248 & $-0.51^{+ 0.48}_{- 0.46}$ & $37.71^{+ 7.77}_{- 6.56}$ & $  9.511/ 16$ & $ 3.99^{+ 1.28}_{- 1.53}$ & $14.26^{+ 4.56}_{- 3.66}$ & $   0.38^{+   0.12}_{-   0.13}$ & $   0.57^{+   0.14}_{-   0.15}$ & $  11.66^{+  61.30}_{-   7.26}$ & $   0.11^{+   0.18}_{-   0.08}$ & $ 10.539/ 15$ & $   0.69^{+   0.09}_{-   0.11}$ & $   0.20^{+   0.01}_{-   0.07}$ & 0.2544 \\
41 & 19:29:41.189 & 0.432 & $-0.54^{+ 0.12}_{- 0.12}$ & $54.94^{+ 1.73}_{- 1.68}$ & $204.166/160$ & $ 4.19^{+ 0.37}_{- 0.34}$ & $16.85^{+ 0.58}_{- 0.55}$ & $   1.07^{+   0.07}_{-   0.07}$ & $   2.86^{+   0.10}_{-   0.10}$ & $  26.87^{+  11.23}_{-   7.64}$ & $   0.28^{+   0.04}_{-   0.04}$ & $189.646/159$ & $   5.23^{+   0.13}_{-   0.13}$ & $   0.32^{+   0.03}_{-   0.03}$ & 1.0000\\
42 & 19:29:42.941 & 0.120 & $-0.06^{+ 0.44}_{- 0.39}$ & $66.48^{+ 6.06}_{- 5.20}$ & $ 27.940/ 33$ & $ 5.42^{+ 1.76}_{- 1.51}$ & $19.29^{+ 2.09}_{- 1.68}$ & $   0.45^{+   0.16}_{-   0.13}$ & $   1.75^{+   0.18}_{-   0.20}$ & $   4.05^{+   9.94}_{-   2.49}$ & $   0.10^{+   0.05}_{-   0.04}$ & $ 24.662/ 32$ & $   0.85^{+   0.05}_{-   0.06}$ & $   0.12^{+   0.01}_{-   0.04}$ & 0.9313 \\

\enddata
\tablenotetext{a}{Events $01-40$ were happened on 2009 January 22, while bursts 41 and 42 were on 2009 January 30.}
\tablenotetext{b}{The luminocity of Blackbody components in units of $10^{39}$\,erg\,s$^{-1}$}
\tablenotetext{c}{Butst energy fluence in GBM band ($8-200$\,keV) in units of $10^{-7}$\,erg\,cm$^{-2}$}
\tablenotetext{d}{Butst energy fluence in XRT band ($0.5-10$\,keV) in units of $10^{-7}$\,erg\,cm$^{-2}$}
\end{deluxetable}

\end{document}